\documentclass[english,11pt,oneside]{article}

\usepackage{amsmath}
\usepackage{amssymb}
\usepackage{cite}
\usepackage{graphicx}
\usepackage{bbm}
\usepackage{color}
\usepackage{url}
\usepackage{authblk}
\usepackage{tikz}
\usetikzlibrary{tikzmark}
\usepackage{multirow}
\usepackage[compat=1.1.0]{tikz-feynman}
\usepackage{graphicx}
\usepackage{hyperref}
\usepackage{multirow}
\usepackage[table,xcdraw]{xcolor}
\usepackage{enumerate}

\usepackage[normalem]{ulem}
\usepackage{comment}
\usepackage{fancyhdr}

\newcommand{\be}{\begin{equation}}  
\newcommand{\ee}{\end{equation}}

\newcommand{\vev}[1]{\langle #1 \rangle}

\newcommand{\SU}[1]{\ensuremath{\mathrm{SU}(#1)}}
\newcommand{\U}[1]{\ensuremath{\mathrm{U}(#1)}}
\newcommand{\into}{\ensuremath{\,\rightarrow\,}}

\newcommand{\MP}{\ensuremath{M_\mathrm{Pl}}}

\definecolor{deepfuchsia}{rgb}{0.76, 0.33, 0.76}

\begin{document}

\pagestyle{fancy}
\fancyhf{}
\fancyhead[R]{ULB-TH/26-22}
\renewcommand{\headrulewidth}{0pt}

\title{\bf The price for monopole dark matter}
\author[1]{Felix Br\"ummer}
\author[2]{Giacomo Ferrante}
\author[1]{Théodore Fischer}
\author[3]{Michele Frigerio}
\affil[1]{\emph{Laboratoire Univers et Particules de Montpellier (LUPM), University of Montpellier \& CNRS, Montpellier, France}}
\affil[2]{\emph{Service de Physique Théorique, Université Libre de Bruxelles (ULB), Brussels, Belgium}}
\affil[3]{\emph{Laboratoire de Physique Théorique et Hautes Énergies (LPTHE), CNRS \& Sorbonne
Université, 4 Place Jussieu, Paris, France}}

\maketitle
\thispagestyle{fancy} 
\pagestyle{plain}

\begin{abstract}
\noindent 
We construct an explicit model where dark matter consists of 't Hooft-Polyakov monopoles. The dark sector is in thermal contact with the Standard Model, and dark monopoles are created by a thermal phase transition in the early Universe. Generically, the abundance of monopoles is negligible with respect to that of stable dark elementary particles. 
We show how to avoid this by taking the lightest stable particle, a dark fermion, sufficiently light for its abundance to be suppressed,  yet heavy enough to satisfy constraints on dark radiation. In this specific window of parameters,
dark matter  is composed of monopoles with a mass of about $10^8$ GeV or larger, depending on the nature of the phase transition. 
This candidate lies beyond the reach of present, conventional dark-matter detection experiments.
However, the model necessarily predicts dark radiation, with $\Delta N_{\rm eff}$ close to present-day bounds. In addition,
if the dark phase transition is strongly first order, we find that the corresponding gravitational wave spectrum lies close to the region probed by future interferometers.
\end{abstract}

\newpage
\tableofcontents

\section{Introduction}

Is dark matter composed of elementary particles? That is, does the majority of what we observe as dark matter consist of quanta of elementary quantum fields, with no substructure up to the scale of post-inflationary reheating? And if not, what else could it be?

Among the alternative scenarios which have been explored, several are nowadays classic: bound states of a strongly coupled dark sector, coherent scalar field oscillations, primordial black holes. But there is a distinct scenario which has seen relatively little attention: dark matter could consist of topological defects, and its stability could be accounted for by its topological charge. Specifically, it could consist of magnetic monopoles of a dark gauge group \cite{Murayama:2009nj}.

In fact, it turns out to be quite hard to construct a model of dark monopole dark matter which is both realistic and calculable. In \cite{Brummer:2025inh} the minimal 't Hooft-Polyakov dark monopole was studied with the following assumptions: monopoles are generated by a thermal dark-sector phase transition, taking place at a temperature where the dark sector is in equilibrium with the Standard Model due to a Higgs portal coupling;  the dark-sector phase transition is independent of the electroweak phase transition, as the former occurs at a much higher scale. Under these assumptions, one finds that the dark monopole abundance is always negligible with respect to the abundance of massive dark-sector gauge bosons, which are stable as they carry a dark electric charge. Remarkably, this conclusion is independent of the nature of the phase transition.

In the present paper, we propose a non-minimal extension of the model which does allow for dark monopoles as the dominant component of dark matter. Dark gauge bosons are now allowed to decay into dark fermions, whose mass is controlled by a different parameter, and whose relic abundance can therefore be suppressed. However, the addition of dark fermions makes the model susceptible to additional constraints: notably, it predicts a sizeable fraction of energy density in dark radiation. As we will show, phenomenological constraints force the model into a fairly narrow corner of its parameter space, and future experiments may well rule it out: monopole dark matter is possible, but it comes at a steep price.
Our analysis will make clear that this conclusion is robust under model variations or extensions.

\section{Dark monopoles, dark gauge bosons and dark fermions}
\subsection{The model}\label{sec:Model}

We consider a dark sector with gauge group $\SU{2}$ and a real scalar triplet $\phi$. The tree-level potential is
\begin{equation}\label{eq:Pot}
    V_{\rm tree}(\phi) = -\frac{m^2}{2} \phi^2 + \frac{\lambda}{4} (\phi^2)^2+\frac{\lambda_{\phi H}}{2}\,\phi^2\,|H|^2\,,
\end{equation}
where $H$ is the Standard Model (SM) Higgs doublet. For $\lambda>0$ and $m^2>0$, and neglecting the Higgs portal term, $V_{\rm tree}$ is minimized at $\vev{\phi^2}=\eta^2$, where
\begin{equation}\label{eq:vev}
\eta^2=\frac{m^2}{\lambda}\,.
\end{equation}
In the minimum, $\SU{2}$ is broken to $\U{1}$. The particle spectrum consists of a scalar radial mode $\rho$ with mass $m_\rho^2 = 2\lambda \eta^2$ and two massive gauge bosons ${W'}^\pm$ with masses $m_{W'}^2 = g^2 \eta^2$, where $g$ is the dark gauge coupling. A third vector boson $\gamma'$, associated to the unbroken $\U{1}$, remains massless.

This model is well known to admit stable magnetic monopole configurations \cite{tHooft:1974kcl, Polyakov:1974ek}.
The lightest monopole $M$ with unit winding number has mass $m_M = k\times 4\pi \eta/g$, where $k=k(\lambda/g^2)$ is a monotonously growing function, with $1\leq k\lesssim 1.8$.
It has magnetic charge $q_M = 4\pi/g$, no electric charge (for minimality, we assume a CP-conserving dark sector, with vanishing theta term for the dark gauge fields), and spin zero.  The monopole core radius $r_M$ scales as $r_M \approx (g \eta)^{-1}$.
The monopole behaves as a classical object since $r_M$ is much larger than its Compton wavelength $\lambda_M= 1/m_M$ for perturbatively small $g$.

As the lightest magnetically charged state in the theory, the monopole is stable, and the same is true for the lightest electrically charged particle $W'$. Both are therefore possible dark matter (DM) candidates. A detailed study of the cosmology of this minimal model \cite{Brummer:2025inh} has shown that, under fairly general assumptions, the DM relic density is always dominated by $W'$ (for a later study see \cite{Xie:2026vor}).

Can the DM relic density instead be dominated by monopoles in a non-minimal model?
With this question in mind, we extend the model by
 introducing two left-handed Weyl fermions $\psi,\chi$ transforming as $\SU{2}$ doublets.\footnote{The somewhat more minimal choice of adding a single fermion doublet is plagued by the presence of the Witten anomaly \cite{Witten:1982fp}. One could replace the fermions with additional scalars, but this would complicate the analysis of the scalar potential and of the vacuum manifold.} 
The fermionic part of the Lagrangian density reads
\be\label{eq:Lferm}
    \mathcal{L} = i \chi^\dagger \bar{\sigma}^\mu D_\mu \chi + i \psi^\dagger \bar{\sigma}^\mu D_\mu \psi - \left[ y\, \chi^i \epsilon_{ij}\phi^a\left(\sigma^a\right)^j_k\psi^k+ m_f \,\chi^i\epsilon_{ij}\psi^j + {\rm h.c.}\right]\,,
\ee
where $D_\mu = \partial_\mu - i\, g W_\mu^a\, \sigma^a/2$ is the covariant derivative, and $\sigma^a$ are the Pauli matrices. This is the most general Lagrangian allowed by a global $\U{1}_F$ symmetry under which $\psi$ and $\chi$ have opposite charges; the effect of $\U{1}_F$ breaking will be discussed shortly.
We once again assume CP conservation, with $y$ and $m_f$ both real. 
After spontaneous symmetry breaking, $\psi$ and $\chi$ arrange themselves into two Dirac fermions $e'$ and $\mu'$ with masses $m_{e'} =- m_f+y\eta$ and $m_{\mu'} = m_f +y\eta$. 
We define $e'$ and $\mu'$ to carry the same dark electric charge, $q=-1/2$. In this convention, they carry opposite $\U{1}_F$ charge and, in particular, $W'$ couples to $e'\mu'$ but not to $e'e'$ nor $\mu'\mu'$.
Therefore, in the exact $\U{1}_F$ limit, the two lightest particles among $W',e',$ and $\mu'$ are both stable. 

Once the assumption of exact $\U{1}_F$ symmetry is relaxed, one can write additional Yukawa couplings $\chi\chi\phi$ and $\psi\psi\phi$; as a consequence, $W'$ couples also to  $e'e'$ and $\mu'\mu'$, and only the lightest particle carrying a dark electric charge is absolutely stable. Without loss of generality it can be taken to be $e'$.
The phenomenologically most interesting scenario will turn out to be
$m_{W'},m_{\mu'}\gg m_{e'}$, with efficient decays of $W'$ and $\mu'$ into stable dark electrons. For simplicity, we assume the $\U{1}_F$-breaking couplings to be small enough not to significantly modify the above expressions for $m_{e'}$ and $m_{\mu'}$.

In the presence of fermions, the monopole background configuration is modified. In the limit of CP conservation, exact $\U{1}_F$, and vanishing Dirac mass $m_f=0$, the monopole background admits a fermionic zero mode, which implies that the monopole ground state is a doublet formed by two degenerate states with fractional fermion charge $\pm 1/2$ \cite{Jackiw:1975fn} and dark electric charge $\pm 1/4$ \cite{Callan:1982au}. This remains the case as long as $|m_f|<|y\eta|$, while for $|m_f|>|y\eta|$ one recovers a non-degenerate monopole ground state, with zero fermion number and zero electric charge \cite{Harvey:1983tp}.\footnote{
How the degeneracy is lifted, and how the fermion and electric charge of the ground state should be defined, is subject to debate. Refs.~\cite{Niemi:1984dq,Niemi:1984vz} finds that these charges vary continuously with $m_f$, vanishing only in the limit $m_f\to \infty$.
For some related, more recent considerations see \cite{Hook:2024als}.}
We will assume the latter situation for simplicity.
In any case, these fermionic effects do not affect the topological stability of the monopole. We assume that the fermion-induced (quantum) deformations of the (classical) monopole mass and shape are small, and that $m_f$ is large enough for the excited monopole states to relax efficiently to the ground state, so that the latter only has cosmological relevance.

We close the discussion of the model by returning to the scalar potential. In general, radiative corrections are small compared to the tree-level potential $V_{\rm tree}$ of Eq.~\eqref{eq:Pot}, and they can be neglected for all practical purposes. An exception is the limit $m^2\rightarrow0$, in which the vacuum structure of the theory is determined radiatively \cite{Coleman:1973jx}. Since we will use this limit to exemplify the case of a strongly first-order phase transition in Sec.~\ref{sec:sFOPT}, we will now briefly discuss it.

We assume that $\lambda \ll g^2$ and thus neglect scalar loops. The one-loop effective potential  in the $\overline{\rm MS}$ renormalisation scheme for a classical background field, also called $\phi$ by a slight abuse of notation, reads
\begin{equation}
    \begin{split}
    V_\text{eff}(\phi)=&-\frac{m^2}{2}\phi^2+\frac{\lambda}{4}(\phi^2)^2
    +\frac{6m_{W'}^4(\phi)}{64\pi^2}\left(\log \frac{m_{W'}^2(\phi)}{\mu^2}-\frac{5}{6}\right)\\
    &-\frac{4m_{\mu'}^4(\phi)}{64\pi^2}\left(\log \frac{m_{\mu'}^2(\phi)}{\mu^2}-\frac{3}{2}\right)
    -\frac{4m_{e'}^4(\phi)}{64\pi^2}\left(\log \frac{m_{e'}^2(\phi)}{\mu^2}-\frac{3}{2}\right).
\end{split}\label{eq:V1loop}
\end{equation}
Here $\mu$ is the renormalisation scale, and the field dependent masses are
\begin{equation}\label{eq:masses}
m_{W'}^2(\phi) = g^2\phi^2\,,\qquad 
m_{\mu'}^2(\phi) = (m_f+y\phi)^2\,,\qquad
m_{e'}^2(\phi) = (m_f-y\phi)^2\,.
\end{equation}
The minimisation of the effective potential allows to express the self-coupling $\lambda$ as a function of the VEV $\eta$ of $\phi$ and the other couplings:
\begin{equation}\label{eq:CWrelation}
    \lambda_0=\frac{m_0^2}{\eta^2}+\frac{1}{8\pi^2}\left[g_0^4+2y_0\frac{m_{\mu',0}^3}{\eta^3}\left(\log\frac{m_{\mu',0}^2}{m_{W',0}^2}-1\right)-2y_0\frac{m_{e',0}^3}{\eta^3}\left(\log\frac{m_{e',0}^2}{m_{W',0}^2}-1\right)\right],
\end{equation}
where the subscript ``0'' indicates that the couplings are evaluated at a fixed reference renormalisation scale $\mu=m_{W',0}=g_0\eta$. 
Eq.~\eqref{eq:CWrelation} generalises the tree-level relation of Eq.~\eqref{eq:vev}.
As $m_0^2\into 0$, the quartic coupling $\lambda$ is indeed loop-suppressed, and so scalar loop contributions to $V_{\rm eff}$ can be self-consistently neglected. The running of the couplings with $\mu$ is discussed in App.~\ref{app:betaFunc}.

\subsection{A sketch of the dark sector cosmology}\label{sec:cosmo}

At very early times and large temperatures above the dark-sector symmetry breaking scale, the visible and the dark sector are in thermal contact via the Higgs-portal coupling $\lambda_{\phi H}$. We take the post-inflationary reheating temperature to be large enough for the dark-sector $\SU{2}$ gauge symmetry to be restored by thermal effects.

The dark sector undergoes a phase transition as the Universe cools, leading to the production of dark monopoles. Depending on their initial abundance, monopole-antimonopole pairs may annihilate efficiently for a while, thus reducing the monopole density. The remaining monopoles form a component of DM. Their abundance will be calculated in Sec.~\ref{sec:mono}.

Dark gauge bosons become massive and decay into $e'$ and $\mu'$ when the temperature drops below their mass. The other unstable dark-sector particles decay as well. One is left with a thermal bath containing $e'$ and dark photons. Thermal equilibrium between the $e'$ and the visible sector thermal bath is lost below a decoupling temperature $T_{\rm dec}$, but the dark electrons continue forming a separate thermal bath with the dark photons until they finally undergo thermal freeze-out, at $T= T_{\rm fo}$, and contribute to the DM relic density. The $s$-wave approximation to the thermally averaged cross-section is
\be
    \langle \sigma v\rangle_{\bar e' e' \rightarrow \gamma'\gamma'} \approx \frac{g^4}{256\pi m_{e'}^2}\,,
\ee
leading to a non-relativistic freeze-out with
\be\label{eq:FermAbun}
    \Omega_{e'}h^2 \approx  0.1 \left(\frac{11}{\gamma_\star(T_{\rm fo})}\right)^{1/2} \left(\frac{z_{\rm fo}}{20}\right)\left(\frac{0.4}{g}\right)^4 \left(\frac{m_{e'}}{100\,{\rm GeV}}\right)^2
\ee
where $z_{\rm fo}\equiv m_{e'}/T_{\rm fo}$, $\gamma_\star \equiv \pi^2 g_\star/90$, and $g_\star$ is the effective number of relativistic degrees of freedom. Since our goal is to construct a model of monopole DM, the dark fermions should contribute a negligible fraction of the observed DM abundance, which requires $m_e'\ll 100 $ GeV. The dark photons remain as dark radiation, with the ensuing constraints discussed in Sec.~\ref{sec:DeltaNeff}.

In the remainder of this paper, we will study the different stages of the cosmological evolution of the dark sector in detail, and compare with the relevant observational constraints.

\section{Dark radiation}
\label{sec:DeltaNeff}

At temperatures below electron-positron annihilation, $T\lesssim 0.5\;{\rm MeV}$, the fraction of energy density of all the relativistic species, relative to the one of the SM photon, is parameterised by the \textit{effective number of neutrino species},
\be
    N_{\rm eff} \equiv \frac{8}{7}\left(\frac{11}{4}\right)^{4/3} \frac{\rho_\nu + \rho_D}{\rho_\gamma}\equiv N_\nu + \Delta N_{\rm eff}.
\ee
In the second equality above, we have defined $\Delta N_{\rm eff}$ to be the contribution coming from dark-sector radiation
\be\label{eq:DeltaNeff}
\begin{aligned}
    \Delta N_{\rm eff}  = \frac{4}{7}\left(\frac{11}{4}\right)^{4/3} 
    \left(\frac{T^{D}_{\rm CMB}}{T^{\rm SM}_{\rm CMB}}\right)^4 
    \left(g^D_B+\dfrac78 g^D_F \right),
\end{aligned}
\ee
where $g^D_B$ and $g^D_F$ are the number of bosonic and fermionic degrees of freedom contributing to dark radiation,
while $T^{D}_{\rm CMB}$ and $T^{\rm SM}_{\rm CMB}$ are the temperatures of the dark sector and SM thermal baths at the time of CMB, respectively. Upon assuming that entropy is separately conserved in each sector, we have
\be
    \frac{T^D_{\rm CMB}}{T^{\rm SM}_{\rm CMB}}=\left[\frac{g_{s}^D(T_{\rm dec})}{g_{s}^D(T^D_{\rm CMB})}\right]^{1/3}
    \left[\frac{g_{s}^{\rm SM}(T^{\rm SM}_{\rm CMB})}{g_{s}^{\rm SM}(T_{\rm dec})}\right]^{1/3},
\ee
where $g_{s}^{\rm SM}$ ($g_{s}^{D}$) are the relativistic degrees of freedom in entropy in the visible (dark) sector, and $T_{\rm dec}$ is the temperature at which the dark sector loses thermal contact with the SM thermal bath (see Sec.~\ref{sec:cosmo}). 
In the absence of any extra radiation, the SM prediction is $N_{\rm eff} = N_\nu=3.044$ \cite{Akita:2020szl,Froustey:2020mcq,Bennett:2020zkv}.

The presence of dark radiation affects the expansion of the Universe, thus modifying its evolution during two key moments: the emission of the cosmic microwave background (CMB) and the time of big-bang nucleosynthesis (BBN). A faster expansion of the Universe results into a suppression of the CMB power spectrum at small scales, and it alters the formation of light elements at BBN. The latest CMB measurements from the South Pole Telescope (SPT), in combination with Planck \cite{Planck:2018vyg} and the Atacama Cosmology Telescope (ACT) \cite{ AtacamaCosmologyTelescope:2025nti}, have tightly constrained the presence of extra radiation, setting $N_{\rm eff} = 2.96 ^{+0.21}_{-0.24}$ \cite{SPT-3G:2025bzu}. However, Ref.~\cite{Tristram:2025you} carried out a separate joint analysis with a more consistent modelling of the foreground, finding $N_{\rm eff} = 3.031 \pm 0.138$. On the BBN side, the Large Binocular Telescope (LBT) has recently measured the abundance of primordial helium with an unprecedented accuracy \cite{Skillman:2026ltj}, leading to the most stringent constraint on extra radiation from BBN. The tightest bound on $\Delta N_{\rm eff}$ comes from a combination of BBN and CMB observations. Such a bound, however, is affected by theoretical uncertainties on the network of reactions taking place during BBN, and, consequently, on the predicted abundance of primordial deuterium. The choice of reaction network has a sizeable impact on the constraint \cite{Giovanetti:2024eff}. In spite of the level of accuracy reached by the latest CMB and BBN observations, constraints on extra radiation are still affected, to some extent, by theoretical uncertainties. Therefore, we use two reference values here: the one given by the Planck collaboration \cite{Planck:2018vyg}, $\Delta N_{\rm eff} <0.3$ at $95\%$ CL, which we regard as the most solid constraint, and the value given by a BBN+CMB analysis after the latest measurement of the $^4\rm He$ abundance, $\Delta N_{\rm eff}<0.125$ at $95\%$ CL \cite{Yeh:2026pil}, keeping in mind the different uncertainties that may affect this number.\footnote{ Notice that the quoted bound from a BBN+CMB joint analysis assumes standard thermodynamical evolution of the Universe between the BBN and CMB times. In other words, $\Delta N_{\rm eff}<0.125$ only applies if $T_{\rm dec} > T^{\rm SM}_{\rm BBN}$, and there are no dark degrees of freedom injecting their entropy between BBN and CMB. In all the other cases, while a dedicated analysis has to be carried out, one expects the predicted value of $\Delta N_{\rm eff}$ to be greatly enhanced, conflicting with constraints from CMB only.}

If, beside the dark photon, the dark electron were to contribute to dark radiation today, then $g^D =2+7/2$, and
the model would be excluded by the $\Delta N_{\rm eff}$ constraint from Planck. For this reason, we will assume that the dark fermions are heavy enough to guarantee $g^D_F=0$.
However, the fermions can still affect $\Delta N_{\rm eff}$ by injecting entropy in the dark sector after $T_{\rm dec}$,
increasing the ratio $T^D_{\rm CMB}/T^{\rm SM}_{\rm CMB}$ and thus $\Delta N_{\rm eff}$.

\begin{figure}[h!]
\begin{center}
\includegraphics[width=.48\textwidth]{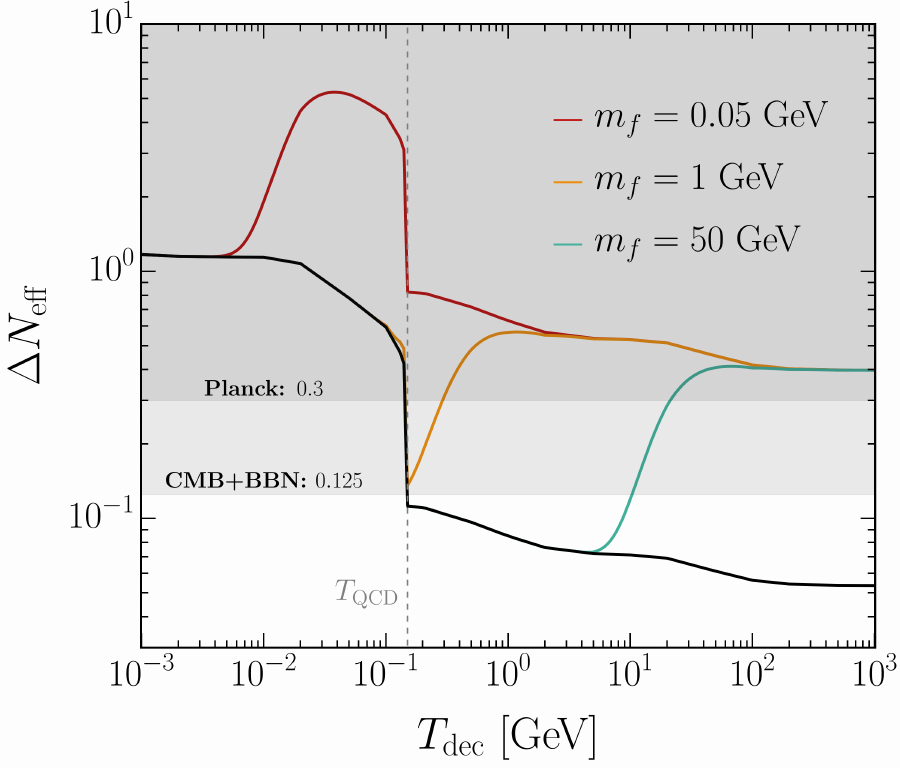}\quad \includegraphics[width=.48\textwidth]{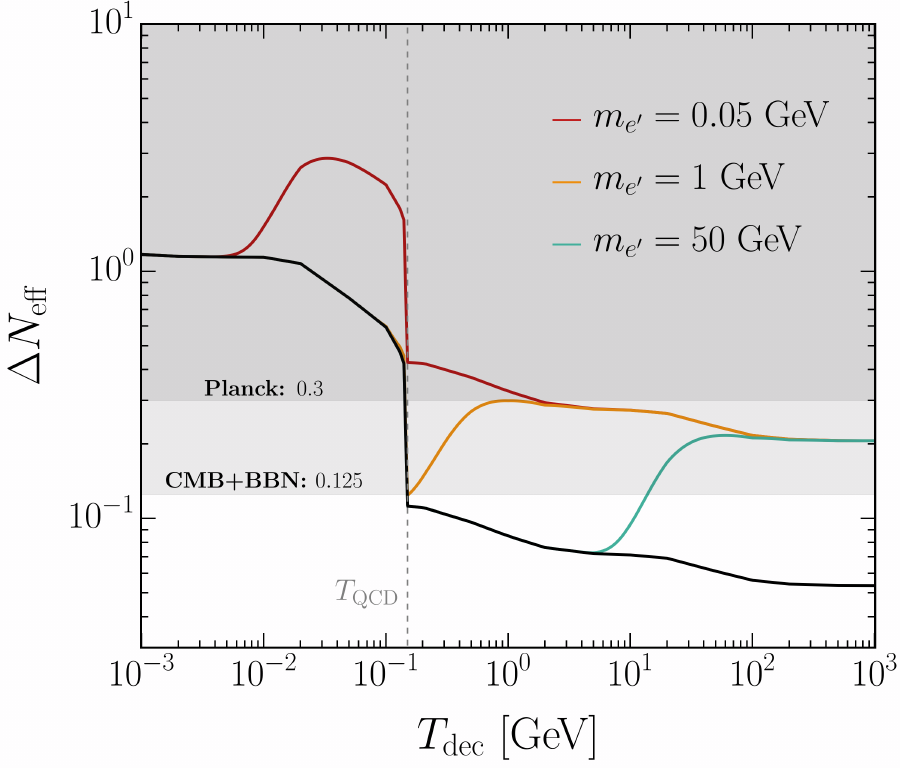}
    \caption{
    The amount of dark radiation, $\Delta N_{\rm eff}$, as a function of the temperature at which the dark sector decouples from the visible one, $T_{\rm dec}$, for different values of the dark fermion masses. In the \textit{left panel} we consider degenerate dark fermions, $m_{e'} = m_{\mu'}=m_f$, while in the \textit{right panel} we take $m_{e'}\ll m_{\mu'}$ with $T_{\rm dec}\ll m_{\mu'}$. The black line illustrates the contribution to $\Delta N_{\rm eff}$ given by the dark photon only.
    The colored lines exhibit a sharp increase at $T_{\rm dec} \sim m_{e'}$: if the two sectors decouple at $T_{\rm dec}>m_{e'}$, the entropy injected by the dark fermions increases the temperature of the dark sector only, thus resulting in a larger value for $\Delta N_{\rm eff}$. Analogously, the drop in the lines at $T_{\rm dec} = T_{\rm QCD}$ corresponds to the decrease of the SM degrees of freedom roughly by a factor of 3 at the QCD phase transition (we used \cite{Husdal:2016haj} to count the SM effective degrees of freedom). The dark (light) gray region is excluded at 95\% CL by the Planck collaboration \cite{Planck:2018vyg} (by a combination of Planck, ACT, SPT and BBN measurements \cite{Yeh:2026pil}).}
\label{fig:Neff}  
\end{center}
\end{figure}

The temperature $T_{\rm dec}$ of thermal decoupling between the dark sector and the SM is estimated as follows. The main process that  maintains thermal contact at low temperatures is $h\leftrightarrow e'\bar{e}'$, with $h$ the SM Higgs boson (see App.~\ref{app:ThermHist} for details). 
The equilibrium condition for this process is given by
\be\label{eq:Tdec}
  \frac{y^2\sin^2\theta}{16\pi}\frac{m_h^3}{m_{e'}^2}\left(1-\frac{4 m^2_{e'}}{m^2_h}\right)^{3/2}\frac{K_1(m_h/T)}{K_2(m_{e'}/T)}\geq H\,,\qquad H=\gamma_\star^{1/2}\frac{T^2}{\MP}\,.
\ee
Here $\theta$ is the $\rho-h$ mixing angle defined in Eq.~\eqref{theta}, $H$ is the Hubble rate, $\MP \simeq 2.4 \times 10^{18}\;{\rm GeV}$ is the reduced Planck mass, and $K_1$ and $K_2$ are modified Bessel functions of the second kind.
Correspondingly, $T_{\rm dec}$ is the lowest temperature for which the inequality \eqref{eq:Tdec} is satisfied.

Let us consider first the case of a negligibly small Yukawa coupling. In the limit $y\to0$, the two dark fermions have the same mass $m_{e'} = m_{\mu'} = m_f$.
Their relic density is determined by the freeze-out of their annihilation into dark photons according to Eq.~\eqref{eq:FermAbun}. The left panel of Fig.~\ref{fig:Neff} shows $\Delta N_{\rm eff}$ as a function of $T_{\rm dec}$ for different values of $m_f$, in the case of a negligibly small $y$. At large $T_{\rm dec}$, the dark fermion contribution to $\Delta N_{\rm eff}$ is larger than allowed by the experimental constraints. This conclusion could in principle be avoided if the dark fermions were to inject most of their entropy while the dark and visible sectors are still in equilibrium, i.e.~$T_{\rm dec}\lesssim m_f/5$. 
However, this is not the case in practice at small $y$, since the $h\leftrightarrow e'\bar{e}'$ rate also decreases with $y$, see Eq.~\eqref{eq:Tdec}. A small $y$ implies $T_{\rm dec}\gg m_f$ and, consequently,  $\Delta N_{\rm eff}\approx 0.4$, which is ruled out.

The preferred scenario is therefore that of a sizeable $y$, which splits the masses of the two fermion families, so that a heavy $\mu'$ becomes non-relativistic and injects its entropy into the common thermal bath formed by the visible and dark sectors long before they decouple.
The right panel of Fig.~\ref{fig:Neff} shows $\Delta N_{\rm eff}$ in the presence of only one light dark fermion $e'$. Clearly, the CMB constraint $\Delta N_{\rm eff}<0.3$ can now be evaded, under the sole condition that the two sectors decouple before the QCD phase transition. It is more difficult to evade the stronger BBN constraint $\Delta N_{\rm eff}<0.125$: this requires $m_{e'}\gtrsim 1$ GeV
and $T_{\rm QCD}<T_{\rm dec} \lesssim m_{e'}/3.5$.

We will see that, in some cases, $T_{\rm dec}$ is forced to be larger then the electroweak scale.
In this regime, our model predicts $\Delta N_{\rm eff} \approx 0.2$, corresponding to a dark electron which injects its entropy into dark photons after the decoupling of the two sectors. Such a scenario
can be definitely tested in the near future, as it is presently on the brink of being excluded.\footnote{
It is possibly to slightly reduce $\Delta N_{\rm eff}$, in models where the dark fermion $e'$ is replace by a dark charged scalar $\rho^+$. In this case,
in the large $T_{\rm dec}$ limit, one finds $\Delta N_{\rm eff} \approx 0.135$. 
}

\section{Phase transitions and monopole abundance}\label{sec:mono}
%%%%%%%%%%%%%%%%%%%%%%%%%%%%%%%%%%%%%%%%%%%%%%%%%%%%%%%%%%%%%%%%%%

The abundance of dark monopoles created by a thermal phase transition is computed from the finite-temperature effective potential for the dark scalar field. At the one-loop level, thermal corrections to Eq.~\eqref{eq:V1loop} are given by
\cite{Dolan:1973qd}
\be\label{eq:DeltaVT}
    \Delta V_T(\phi) =\frac{T^4}{2\pi^2}\left[ \sum_b n_b J_b\left(\frac{m_b^2(\phi)}{T^2}\right) - \sum_f n_f J_f\left(\frac{m_f^2(\phi)}{T^2}\right)\right]\,.
\ee
Here $n_b$ ($n_f$) is the number of degrees of freedom of the bosonic (fermionic) particle $b$ ($f$), whose mass in a $\phi$ background is $m_b(\phi)$ ($m_f(\phi)$). The thermal functions $J_b$ and $J_f$ are defined by
\be
    J_{b,f}(x) = {\int_0}^\infty dq\, q^2 \log\left[1\mp \exp\left(-\sqrt{q^2+x}\right)\right].
\ee
In what follows, we will be interested in the high-temperature limit, $m^2_{b,f}(\phi)\ll T^2$, in which the thermal functions can be expanded as
\be
\begin{aligned}\label{eq:highTexp}
J_b(x) &\approx \frac{\pi^4}{45}+ \frac{\pi^2}{12}x -\frac{\pi}{6}x^{3/2}-\frac{x^2}{32} \log \frac{x}{a_b},\\
J_f(x) &\approx \frac{7\pi^4}{360}-\frac{\pi^2}{24}x - \frac{x^2}{32}\log \frac{x}{a_f},
\end{aligned}
\ee
where $a_f = \pi^2 \exp\left(3/2-2\gamma_{E}\right)$, $a_b = 16 a_f$, and $\gamma_E\simeq 0.577$ is the Euler-Mascheroni constant. 

If the temperature at which the Universe reheats after inflation is much larger than the symmetry breaking scale $\eta$, thermal corrections deform the zero-temperature potential, stabilizing the symmetry preserving point, $\phi=0$, which becomes a global minimum: the $\SU{2}$ symmetry is restored by thermal effects. As the Universe expands, the temperature of the thermal bath decreases until, below some critical temperature $T_c$, the origin ceases to be a global, stable minimum, and a phase transition to a symmetry-breaking minimum may take place. This can happen either smoothly, as in a second-order phase transition (SOPT), or via tunnelling across a barrier separating two local minima, as in a first-order phase transition (FOPT). 

The nature of the thermal phase transition depends on the ratio $\lambda/g^2$, as reviewed e.g.~in \cite{Brummer:2025inh}; for $g^2>\lambda$ the transition is first-order, and for $g^2\ll\lambda$ it is second-order. We take the Yukawa coupling $y$, as well as the Higgs portal coupling $\lambda_{\phi H}$, to be small enough not to affect the nature of the phase transition or to destabilize the effective potential. In particular, we impose $y^2\lesssim \max(g^2,\lambda)$.
However, $\lambda_{\phi H}$ is chosen to be large enough for the dark and the visible sectors to be in thermal equilibrium at $T=T_c$. Requiring that $\phi-H$ scatterings are efficient down to the critical temperature gives \cite{Brummer:2025inh}
\be\label{eq:HiggsPortTherm}
    \lambda_{\phi H}^2 \gtrsim   \frac{10^{14}\,{\rm GeV}}{T_c}\,.
\ee

When the phase transition takes place, the Universe is divided into regions of typical size $\xi$, the correlation length of the scalar field. Inside each region, $\phi$ takes a random direction in the vacuum manifold $S^2$. Therefore, monopoles may form at the intersections of the different domains, with a number density $n_M \approx \xi^{-3}/8$ \cite{Kibble:1976sj}. The monopole relic density today is then given by
\be
    \Omega_M h^2 \simeq 0.12 \left(\frac{Y_M}{4.36\times 10^{-18}} \right) \left(\frac{m_M}{10^8\;{\rm GeV}}\right)\,,
\ee
where the monopole yield $Y_M$ is defined in terms of $n_M$ and the entropy density $s$ as
\be
Y_M=\frac{n_M}{s}\,,\qquad s=4\gamma_\star T^3\,.
\ee

As it will be reviewed in  the following subsections, the computation of the correlation length $\xi$ strongly depends on the details of the phase transition. After monopoles have been produced, their number density can be reduced by monopole-antimonopole annihilation \cite{Zeldovich:1978wj,Preskill:1979zi}, which will be discussed in Sec.~\ref{sec:MonoAnn}.

\subsection{Second-order phase transitions}\label{sec:SOPT}
%%%%%%%%%%%%%%%%%%%%%%%%%%%%%%%%%%%%%%%%%%%%

For $g^2<\lambda$, perturbation theory is unreliable around the critical temperature $T_c$, and therefore cannot predict the nature of the phase transition (see e.g.~\cite{Arnold:1992rz}). In the $g\to 0$ limit, however, dimensional reduction \cite{Ginsparg:1980ef} maps the theory to the ${\rm O}(3)$ model in three Euclidean dimensions, where the phase transition is well known to be of the second order \cite{Zinn-Justin:2002ecy}. By continuity, we expect the phase transition to be also of the second order for small, non-zero $g$. 

During a SOPT, monopoles are formed by the Kibble-Zurek mechanism \cite{Kibble:1976sj,Zurek:1985qw,Zurek:1996sj,delCampo:2013nla}. The correlation length formally diverges at the critical temperature, but so does the relaxation time of the system. The resulting effective correlation length which dictates the monopole abundance is 
\be
\label{eq:correlation_lgth}
    \xi \approx H(T_c)^{-1}\, \left[H(T_c)\xi_0\right]^\frac{1}{1+\nu}\,,
\ee
where  $\xi_0^2\approx 1/m_\rho^2 =  1/(2\lambda\eta^2)$ and $\nu\approx 0.7$ is a critical exponent \cite{Zurek:1996sj, Murayama:2009nj}. 
When $y^2,g^2\ll \lambda$, the critical temperature can be estimated as $T_c^2\approx\frac{12}{5}\eta^2$, and the monopole yield becomes \cite{Brummer:2025inh,Murayama:2009nj}
\be
\begin{aligned}\label{eq:MonoSOPT}
Y_M \approx \frac{1}{32} \left(\frac{5}{6}\lambda\right)^{3/(2+2\nu)} \gamma_\star{}^{(\nu-2)/(2+2\nu)}\left(\frac{T_c}{\MP}\right)^{3\nu/(1+\nu)}\,.
\end{aligned}
\ee

\begin{figure}[ht]
\begin{center}
\includegraphics[width=.55\textwidth]{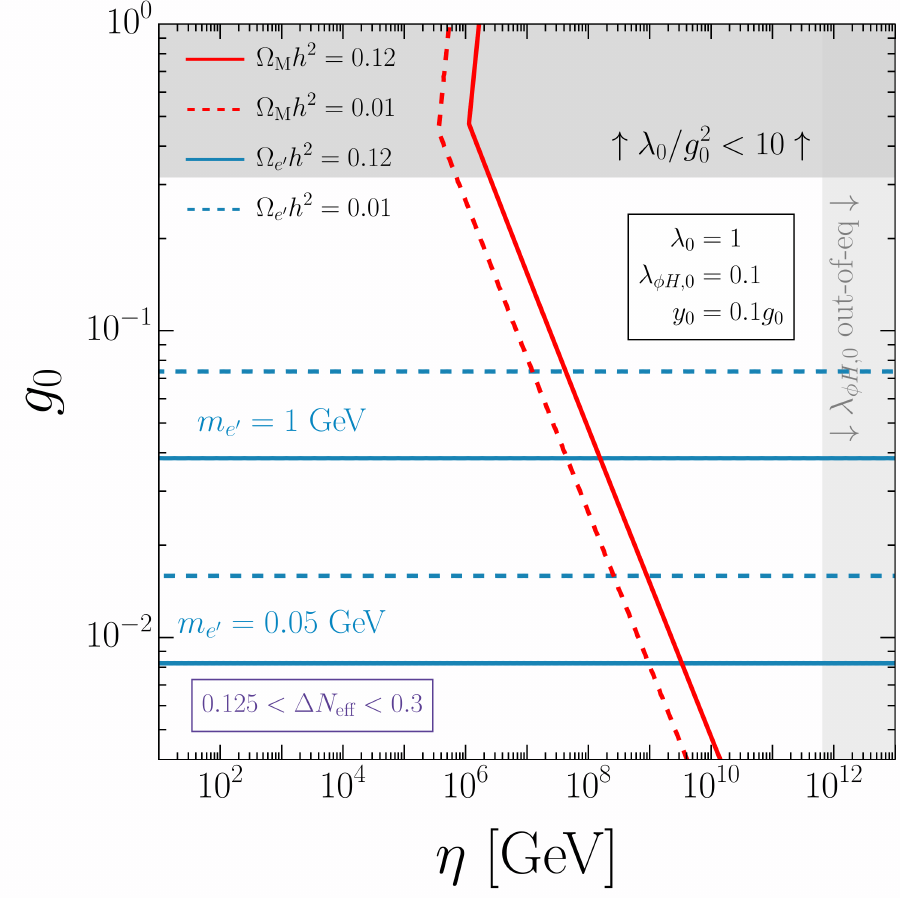}
    \caption{Relic density of the dark electrons $e'$ (blue) and of the dark monopoles $M$ (red) after a SOPT, as a function of the  dark symmetry breaking scale $\eta$ and the dark gauge coupling $g$. The other couplings are fixed as indicated in the legend, and the subscripts zero stand for a reference renormalisation scale, $\mu_0 = m_{W',0}=g_0\eta$. Two representative values of $m_{e'}$ are displayed, corresponding to an appropriate choice of the fermion mass parameter $m_f$.
    The slope of the monopole line for $g_0\gtrsim 0.5$ ($g_0\lesssim 0.5$) is determined by the Kibble-Zurek mechanism (by subsequent monopole-antimonopole annihilations).
    In the light gray region, the Higgs portal is not sufficient to thermalise the dark sector with the SM before the critical temperature. Above some critical ratio $g_0^2/\lambda_0\sim{\cal O}(0.1)$, one expects the phase transition to become first order, as indicated by the dark gray region. 
    In the whole plane, the amount of dark radiation lies within the range indicated in purple.
    }
\label{fig:SOPTFermi}  
\end{center}
\end{figure}

We emphasize again that the monopoles produced by the Kibble-Zurek mechanism can undergo later annihilations, see Sec.~\ref{sec:MonoAnn}. The overall resulting monopole relic density for a SOPT is shown in Fig.~\ref{fig:SOPTFermi}, along with the dark electron relic density for two different values of $m_{e'}$.

The above discussion is valid as long as the monopoles are effectively point-like: $r_M\sim(g\eta)^{-1}<\xi$. We find that this condition is always satisfied in the phenomenologically relevant part of parameter space: a relatively large $\eta$ is required for a significant monopole abundance, and a relatively large $g$ is needed to suppress the $e'$ abundance, see Eq.~\eqref{eq:FermAbun}.

\subsection{First-order phase transitions}\label{sec:MonoFOPT}
FOPTs proceed via tunnelling across a potential barrier in the thermal effective potential, with bubbles of true vacuum nucleating in a background of false vacuum. Considering only the effect of thermal fluctuations, i.e.~neglecting quantum tunnelling, the nucleation rate per unit volume is given by \cite{Linde:1981zj}
\be\label{eq:Gammav}
    \Gamma(T) \approx T^4 \left(\frac{S_3/T}{2\pi}\right)^{3/2}e^{-S_3/T},
\ee
where $S_3$ is the Euclidean action of a bubble, evaluated along the ${\rm O}(3)$-symmetric bounce solution which describes finite-temperature transitions.

The tunnelling process becomes efficient at the nucleation temperature $T_n$, defined by
\be\label{eq:Tn}
    \Gamma(T_n) \approx H(T_n)^4,
\ee
which marks the onset of the phase transition. If the tunnelling rate is sufficiently suppressed, the Universe remains stuck in the false vacuum for a long time before the first bubbles nucleate. Eventually, its vacuum energy density may come to dominate over the energy density contained in radiation, starting at a temperature $T_{\rm eq}$ which is implicitly given by
\be
    T_{\rm eq} = \left( \frac{\Delta V(T_{\rm eq})}{3\gamma_\star(T_{\rm eq})}\right)^{1/4}\,.
\ee
If $T_n < T_{\rm eq}$, the phase transition takes place in a vacuum-dominated universe, and it is said to be \textit{supercooled}.

Once nucleated, bubbles of the new phase grow, driven by the  potential energy difference across the bubble wall $\Delta V(T)$. The progress of the phase transition is described by the probability $\mathcal{P}_{f}$ of finding one point in space in the false vacuum, at some given time or temperature \cite{Guth:1979bh}
\be\label{eq:PfT}
    \mathcal{P}_{f}(T) = \exp\left[-\frac{4\pi}{3}\int_{T}^{T_c} dT' \frac{\Gamma(T')}{T'^4 H(T')}\left(\int_{T}^{T'} v_w \frac{dT''}{H(T'')}\right)^3\right]\equiv e^{-I(T)}\,.
\ee
In this expression, $v_w$ is the bubble wall velocity, which is taken to be constant since expanding bubbles are expected to quickly  approach their asymptotic velocity. Moreover, one assumed that the critical radius $R_c$ at which bubbles are nucleated is small with respect to the size to which bubbles have grown at the end of the phase transition.

The transition completes at the temperature $T_*$, when bubbles collide and the entire Universe enters the broken phase. Monopoles are produced with a correlation length given by the average bubble radius at completion, $\xi = R_ *$, and a comoving number density
\be
    Y_M = \frac{1}{32}\left(\gamma_\star^{\rm reh}\right)^{-1}(R_* T_{\rm reh})^{-3}.
\ee
Here $T_{\rm reh}$ is the temperature at which the Universe is reheated after the phase transition. In generic cases, the moment of completion coincides with the percolation temperature $T_p$,  which is the temperature where 29\% of the Universe has been converted to the true vacuum, 
\be\label{eq:Tp}
\mathcal{P}_f(T_p ) = e^{-I(T_p)} =0.71\,. 
\ee
Supercooled transitions may represent an exception with $T_*\neq T_p$ \cite{Levi:2022bzt,Athron:2023xlk}, as we will discuss below. 

We define the quantity $\beta$ by
\be\label{eq:beta}
    \beta \equiv \frac{d\log \Gamma}{d t} = -H(T) T \frac{d \log \Gamma}{d T}.
\ee
In the second equality we have used $dT = -H(T)Tdt$, which holds as long as entropy is conserved. $\beta_* \equiv \beta(T_*)$ represents the inverse duration of the phase transition. In the following, we will dub as \textit{``fast''} those phase transitions for which $\beta_{*}/H(T_*) \gg 1$. 

Finally, we define the latent heat parameter $\alpha$ of the phase transition as the ratio between vacuum and radiation energy densities at $T_*$
\be\label{eq:alpha}
    \alpha \equiv \frac{\Delta V(T_*)}{\rho_r(T_*)} \approx \left(\frac{T_{\rm eq}}{T_*}\right)^4,
\ee
where we have approximated the potential energy difference with its zero-temperature value, $\Delta V(T_*) \approx \Delta V(T_{\rm eq}) \approx \Delta V$. Values of $\alpha \gg 1$ correspond to \textit{strongly first-order phase transitions} (sFOPT); supercooled transitions are typically strongly first-order. On the contrary, \textit{weakly first-order phase transitions} (wFOPTs) are characterised by $\alpha \lesssim 1$.

As we have reviewed, a FOPT is essentially described by the bounce action $S_3$. To compute $S_3$, we use the high-temperature expansion of the thermal effective potential of Eq.~\eqref{eq:highTexp}. The high-temperature expansion may be unsuitable for field values around the true vacuum, but it does capture the leading thermal corrections at field values across the barrier, which is all that is needed for tunnelling, and therefore for the study of the phase transition. Implementing the high-temperature expansion, and neglecting scalar-loop corrections (i.e.~assuming $\lambda\ll g^2$), the potential takes a polynomial form,
\be\label{eq:VPoly}
    V(\phi,T)\approx\frac{m_T^2}{2}\phi^2-\frac{\delta_T}{3}\phi^3+\frac{\lambda_T}{4}\phi^4,
\ee
where, using Eqs.~\eqref{eq:V1loop}, \eqref{eq:DeltaVT} and \eqref{eq:highTexp},
\be\label{eq:ParamsT}
\begin{aligned}
    m_T^2 &= -m^2 + \left(\frac{g^2}{2} + \frac{y^2}{3}\right)T^2 - \frac{3y^2 m_f^2}{2\pi^2}\left(\log\frac{a_f T^2}{\mu^2}-\frac{3}{2}\right),\\
    \delta_T &=  \frac{3g^3}{2\pi}T,\\
    \lambda_T &= \lambda + \frac{6 g^4}{16\pi^2}\left(\log \frac{a_b T^2}{\mu^2} - \frac{5}{6}\right) -\frac{8 y^4}{16\pi^2}\left(\log \frac{a_f T^2}{\mu^2} - \frac{3}{2}\right).
\end{aligned}
\ee
For a quartic potential such as Eq.~\eqref{eq:VPoly}, one can obtain semi-analytic expressions for $S_3$ in terms of a single parameter \cite{Adams:1993zs,Levi:2022bzt,Salvio:2023ynn},
\be\label{eq:S3}
    S_3 = 4\pi Z_\phi^{3/2}\frac{m_T^3}{\delta_T^2} \times \begin{cases}
   &\dfrac{3}{4}\,\dfrac{1+\exp\left(-1/\sqrt{-\kappa_T}\right)}{2/9-\kappa_T},\quad \kappa_T <0\\[12pt]
    &\dfrac{F(\kappa_T)}{6\left(\kappa_T-2/9\right)^2} ,\quad \kappa_T\geq0
    \end{cases}\qquad {\rm with}\qquad \kappa_T \equiv \lambda_T \frac{m_T^2}{\delta_T^2}.
\ee
In contrast with previous studies, we include the dependence of $S_3$ on $Z_\phi$, the scalar-field wavefunction renormalisation defined by Eq.~\eqref{eq:Zphi}, according to the derivation detailed in App.~\ref{app:RGImprov}. The fitting function $F$ appearing in Eq.~\eqref{eq:S3} is \cite{Levi:2022bzt}
\be
\begin{aligned}
  F(x) \simeq \frac{16}{243}\Bigg[&1-38.23\left(x-\frac{2}{9}\right) + 115.26 \left(x-\frac{2}{9}\right)^2 \\&+ 58.07 \sqrt{x}\left(x-\frac{2}{9}\right)^2 + 229.07 x \left(x-\frac{2}{9}\right)^2\Bigg].
\end{aligned}
\ee
The coefficients of the potential in Eq.~\eqref{eq:VPoly} are functions of the couplings, as well as of the temperature, and, as such, they run with the renormalisation scale $\mu$. The $\beta$ functions describing the running of the couplings are given in App.~\ref{app:betaFunc}. The running is particularly relevant in supercooled phase transitions, where there is a large separation of scales. We will comment more on this in Sec.~\ref{sec:sFOPT}.

So far our discussion has been general and applies to any FOPT. We now turn to the specificities of weakly and strongly FOPTs.

\subsubsection{Weakly first-order phase transitions}\label{sec:wFOPT}
%%%%%%%%%%%%%%%%%%%%%%%%%%%%%%%%%%%%%%%%%%%%%%%%%%%%%%%%%%%%%%%%%%%%%%

As already explained, a wFOPT corresponds to $\alpha \lesssim 1$. For concreteness, we consider $\lambda_{\phi H},y^2 \ll g^2$ and $\lambda=0.1 g^2$ as a case study. For this choice of couplings, we checked that $\alpha \ll 1$ for all the relevant values of $g$ and $\eta$.
We also checked that the high-temperature expansion provides a good approximation of the thermal corrections at $T\approx T_c$, for values of $\phi$ as large as the absolute minimum of the effective potential \cite{Dolan:1973qd, Anderson:1991zb, Dine:1992wr, Arnold:1992rz}.

For this parameter choice, the dominant corrections to the finite-temperature potential are due to gauge-boson one-loop diagrams. Neglecting all others, Eq.~\eqref{eq:VPoly} becomes 
\be\label{eq:VwFOPT}
    V(\phi,T) = \frac{1}{2}\left(\frac{g^2}{2}T^2 - \lambda \eta^2+\frac{g^4}{8\pi^2}\eta^2\right)\phi^2 - \frac{g^3}{2\pi}T\phi^3 + \frac{1}{4}\left[\lambda+\frac{6 g^4}{16\pi^2}\left(\log\frac{a_b T^2}{g^2\eta^2}-\frac{5}{6}\right)\right]\phi^4\,,
\ee
where we used Eq.~\ref{eq:CWrelation} (which strictly holds for $\mu=m_{W'}$) to relate $m^2$ to $\eta^2$.
This expression does not include any resummation of leading higher-loop ``daisy'' diagrams \cite{Dolan:1973qd,Parwani:1991gq,Arnold:1992rz}; we have checked that including daisy resummation would have no significant impact on our results. 
The temperature at which the thermal barrier disappears (i.e.~when $m_T^2$ becomes equal to $0$) is 
\be
    T_0=\eta\sqrt{\frac{2\lambda}{g^2}-\frac{g^2}{4\pi^2}}\, ,
\ee
while the critical temperature $T_c$ is computed numerically and found to be close above $T_0$.

\begin{figure}[tb!]
\begin{center}
\includegraphics[width=.48\textwidth]{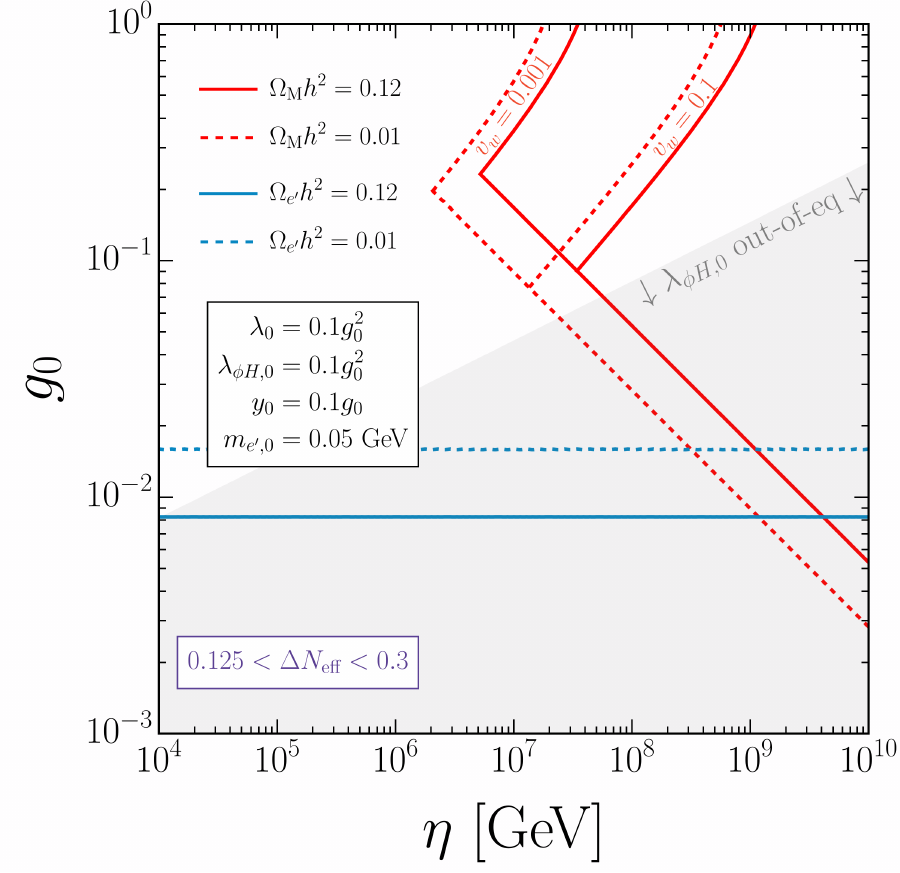}\quad\includegraphics[width=.48\textwidth]{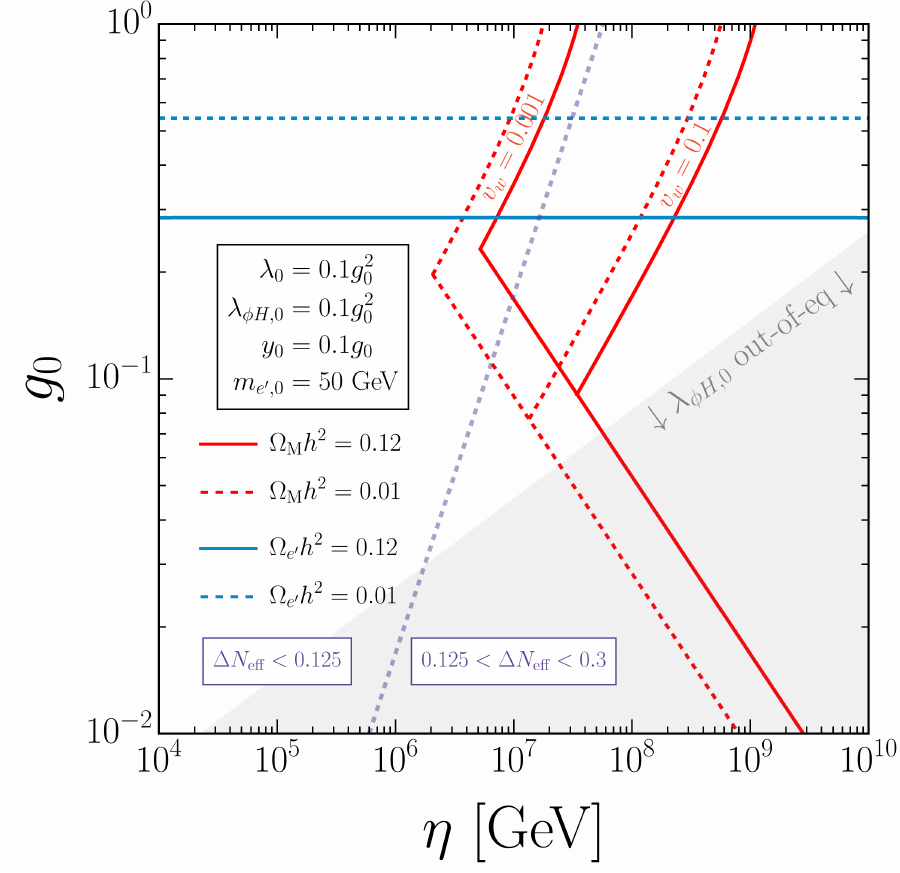}
    \caption{
    Relic density of the dark electrons $e'$ (blue) and of the dark monopoles $M$ (red) after a wFOPT, as a function of the  dark symmetry-breaking scale $\eta$ and the dark gauge coupling $g_0$, with $m_{e'}=0.05\,(50)$ GeV in the left (right) panel.
    The other couplings are fixed as indicated in the legend, and the subscripts zero stand for the reference renormalisation scale, $\mu_0 =g_0\eta$. 
    The slope of the monopole lines for $g_0\gtrsim 0.1$ is determined by the bubble radius at percolation, for two benchmark values of the bubble wall velocity $v_w$. For $g_0\lesssim 0.1$, the final abundance is rather determined by subsequent monopole-antimonopole annihilations.
    In the light gray region, the Higgs portal is not sufficient to thermalise the dark sector with the SM before the critical temperature.
    In the right panel only, the amount of dark radiation can be lower than the combined CMB+BBN, in the region to the left of the purple line (see ranges for $\Delta N_{\rm eff}$ indicated in purple).}   
\label{fig:MonoFermwFOPT}  
\end{center}
\end{figure}

As $T_0$ is smaller than but close to $T_c$, bubbles nucleate and percolate very soon after the critical temperature has been reached. There is no time for the vacuum energy density to come to dominate: the transition is weak, $\alpha< 1$, and fast, $\beta_{*}\gg H(T_*)$. 
Therefore, we can regard the Hubble parameter as constant over the typical time-scale of the transition and  write the exponent in Eq.~\eqref{eq:PfT} as \cite{Hindmarsh:2019phv} 
\be\label{eq:IwFOPT}
    I(T) \approx 8\pi v_w^3 \frac{\Gamma(T)}{\beta(T)^4}\,,
\ee
which allows for a numerical computation of the percolation temperature. Analogously, the average bubble radius at completion can be computed as \cite{Hindmarsh:2019phv}
\be\label{eq:R*wFOPT}
    R_* = \left[\frac{8\pi v_w^3}{1-\mathcal{P}_f(T_{*})}\right]^{1/3}\beta_*^{-1}\,.
\ee

Since for a wFOPT the Universe remains radiation dominated and its evolution approximately adiabatic, the reheating temperature coincides with the percolation one, $T_{\rm reh}\approx T_p$. The last parameter needed to compute the monopole abundance produced by the transition is the bubble wall velocity, cfr.~Eq.~\eqref{eq:IwFOPT} and Eq.~\eqref{eq:R*wFOPT}. On general grounds, we expect the radiation bath to exert a non-negligible friction on bubble walls, so that $v_w\ll 1$. A precise estimate of $v_w$ goes beyond the scope of this work and we treat it here as a free (small) parameter. Fig.~\ref{fig:MonoFermwFOPT} shows the monopole abundance produced by a wFOPT, for two benchmark values of $v_w$.

\subsubsection{Strongly first-order phase transitions}\label{sec:sFOPT}
%%%%%%%%%%%%%%%%%%%%%%%%%%%%%%%%%%%%%%%%%%%%%%%%%%%%%%%%%%%%%%%%%%%%%%%

A sFOPT can be realised, in particular, in classically scale-invariant potentials, where the potential barrier is due to loop effects, such as Eq.~\eqref{eq:V1loop} in the limit $m^2\rightarrow 0$. While for $T<T_c$ the high-temperature expansion is not a good approximation of the potential for field values as large as the true vacuum (see e.g.~\cite{Jaeckel:2016jlh}), it can still be used to describe the region of field space relevant for the tunnelling process. Therefore, for the purpose of computing the bounce action, the thermal effective potential in the classically-conformal model can be written in the form of Eq.~\eqref{eq:VPoly}.

Since in sFOPTs there is typically a large separation between $T_c$ and $T_*$, some comments are in order:

\begin{enumerate}
\item All the couplings appearing in Eq.~\eqref{eq:ParamsT} are to be understood as functions of the renormalisation scale $\mu$. For the zero-temperature effective potential, the loop expansion coincides with an expansion in powers of the couplings, and the effective potential is $\mu$-independent order by order. By contrast, the one-loop corrections at finite temperature carry some residual renormalisation-scale dependence, which is of the order of zero-temperature, one-loop effects (for a pedagogical review, see \cite{Gould:2021oba}). This issue is particularly relevant in sFOPTs when there is a large hierarchy of scales due to supercooling. In order to solve it properly, one needs to include two-loop thermal corrections, either explicitly or via Renormalisation-Group improvement of the one-loop potential. However, it has been argued (in the context of massless scalar QED) that choosing $\mu$ in the one-loop effective potential to be the ``hard'' thermal mass scale of the Matsubara modes, $\mu=\pi T$, reproduces quite well the results of more sophisticated two-loop computations \cite{Kierkla:2023von, Christiansen:2025xhv}. This is the prescription we will adopt in the following.

\item The parameters of the thermal effective potential given in Eq.~\eqref{eq:ParamsT} are obtained neglecting the contributions of scalar loops. This is justified by Eq.~\eqref{eq:CWrelation} which holds at the reference scale $\mu_0=m_{W',0}$, and which implies $\lambda_0 \ll g_0^2, y_0^2$. As the couplings run, Eq.~\eqref{eq:CWrelation} no longer holds at $\mu<\mu_0$, and $\lambda$ may become of the same order as $g^2$. However, we have checked that $\lambda \ll g^2, y^2$ remains true for the range of couplings and temperatures of interest to us.

\item Eq.~\eqref{eq:ParamsT} includes the fermionic contribution to the high-temperature expansion of the thermal effective potential. However, for temperatures and field values such that $T<m_{\mu',e'}(\phi)$, the heavy fermions decouple from the thermal bath. While this behavior is captured by the exact expression for the thermal corrections given in Eq.~\eqref{eq:DeltaVT}, it has to be enforced ``by hand'' at the level of Eq.~\eqref{eq:ParamsT}. The physical fermion masses are given by $m_{\mu',e'}=m_f\pm y\eta$, and  we are interested in the parameter space region with one and only one light fermion $e'$, i.e.~$y\approx m_f/\eta$. The field-dependent fermion masses can thus be written as $m_{\mu',e'}(\phi) \approx m_f(1\pm \phi/\eta)$. Since in supercooled transitions the tunnelling process happens at field values $\phi \ll \eta$ \cite{Cutting:2020nla}, we use for simplicity a step-function decoupling, removing the fermionic contributions from the effective potential for $T< m_f$. Similar remarks apply to the fermion contributions to the beta functions, see App.~\ref{app:betaFunc}.

\end{enumerate}

If the transition takes place during vacuum domination, the function $I$ of Eq.~\eqref{eq:PfT} can be written, for $T_*<T<T_{\rm eq}$, as \cite{Levi:2022bzt} 
\be
    I(T) \approx 8\pi v_w^3 \frac{\Gamma(T)}{\beta(T)^4}\left[1+6\,\frac{H}{\beta(T)}+11\left(\frac{H}{\beta(T)}\right)^2+6 \left(\frac{H}{\beta(T)}\right)^3\right]^{-1}.
\ee
    While the percolation temperature represents a good estimate of $T_*$ in wFOPTs, in sFOPTs the accelerated expansion of the Universe may prevent bubbles from colliding even if the percolation threshold, $\mathcal{P}_f = 0.71$,  has been reached \cite{Athron:2023xlk}. This happens, in particular, in relatively slow transitions, for which the expansion of the Universe is relevant. Due to the exponential growth of the scale factor, $a(t)\approx \exp(Ht)$, the space in the false vacuum separating the bubbles grows sizably during the typical timescale of the phase transition. As a consequence, the physical volume of the false vacuum, $a(t)^3 \mathcal{P}_f$, may still grow for $T<T_p$. 
    In this scenario, the transition can be said to complete as soon as the physical false vacuum fraction starts decreasing,
\be
    \frac{d}{dt}\left[a(t)^3 \mathcal{P}_f(t)\right]<0\,.
\ee
This condition is equivalent to $I(T) > 3 H/\beta(T)$. We define $T_d$ to be the temperature at which the physical volume in the false vacuum starts decreasing,
\be\label{eq:Td}
    T_d:\qquad I(T_d) = 3 \frac{H}{\beta(T_d)}\ .
\ee
Therefore, we set $T_*= \min \left[T_p,T_d\right]$ for sFOPTs. By comparing Eq.~\eqref{eq:Tp} and Eq.~\eqref{eq:Td}, we see that $T_d < T_p$ only for relatively slow supercooled transitions with $\beta_{*}/H \lesssim 9$. 

For transitions taking place during vacuum domination, the average bubble radius at completion can be computed as \cite{Levi:2022bzt}
\be
    R_* = \left\{\frac{\Gamma(T_*)}{\beta_*}I(T_*)^{-1-3\frac{H}{\beta_*}}\left[\tilde{\Gamma}\left(1+3\frac{H}
    {\beta_*},0\right)-\tilde{\Gamma}\left(1+3\frac{H}{\beta_*},I(T_*)\right)\right]\right\}^{-1/3},
\ee
where $\tilde{\Gamma}$ is the incomplete gamma function. This expression typically scales as the duration of the phase transition, $R_* \propto 1/\beta_*$.

\begin{figure}[tb]
\begin{center}
\includegraphics[width=.48\textwidth]{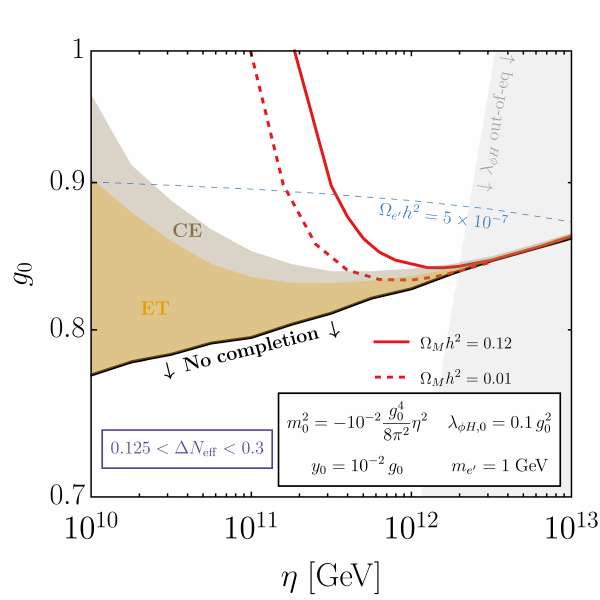}\quad \includegraphics[width=.48\textwidth]{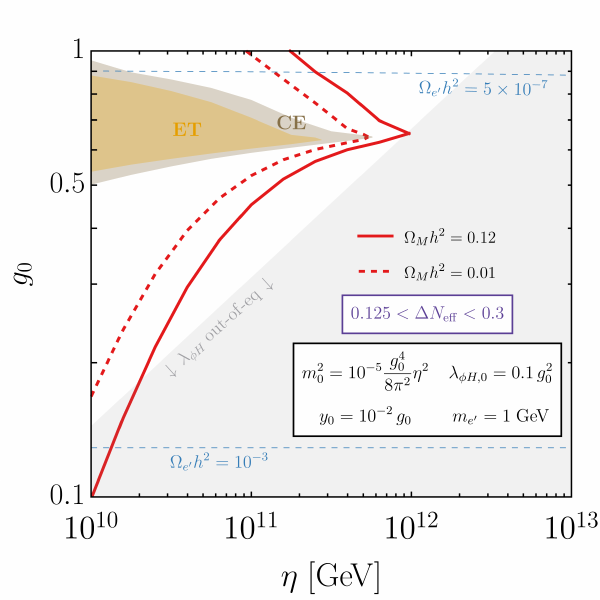}
    \caption{Isocontours of monopole relic density (red) in the scenario of a supercooled phase transition, for a small $m_0^2<0$ (\textit{left panel}) and for a small $m_0^2>0$ (\textit{right panel}). 
    The blue, dashed lines represent the abundance of dark electrons, as given by the freeze-out of their annihilations into dark photons. The ochre and gray shaded regions denote the projected sensitivities of the Einstein Telescope \cite{ET:2019dnz} and the Cosmic Explorer \cite{Reitze:2019iox}, respectively. In the light gray region, the Higgs portal interaction is not efficient enough for the visible and dark sectors to thermalise with each other above the critical temperature, cfr.~Eq.~\eqref{eq:HiggsPortTherm}. In the \textit{left panel}, below the black line, the thermal transition never completes, and the Universe remains stuck in the false vacuum indefinitely, while in the \textit{right panel} the barrier eventually vanishes, due to the positive $m^2_0$. In the whole plane, $\Delta N_{\rm eff}$ lies in the range indicated in purple.
    }
\label{fig:MonoFermsFOPT}  
\end{center}
\end{figure}

The relic abundance of monopoles produced during a sFOPT is shown in Fig.~\ref{fig:MonoFermsFOPT} for a small negative and a small positive value of $m_0^2 \ll \eta^2$.
For $m_0^2<0$ (left panel), the second derivative of the zero-temperature effective potential at the origin is positive, see Eq.~\eqref{eq:V1loop}. A loop-induced barrier separates the origin and the global minimum $\eta$, even at $T=0$. As a result, for $g_0$ sufficiently small ($g_0 \lesssim 0.8$ for the case shown in the left panel of Fig.~\ref{fig:MonoFermsFOPT}), the tunnelling rate is suppressed at low temperatures and the phase transition never completes, with the Universe remaining stuck in the false vacuum indefinitely. On the other hand, for $m_0^2>0$ (right panel) the second derivative of the zero-temperature effective potential at the origin is negative. Therefore, for temperatures $T^2< T_0^2 \approx 2m^2/g^2$, the quadratic coefficient of the thermal effective potential, $m_T^2$, turns negative and the thermal barrier vanishes. As a result, if the transition has not happened yet, it takes place at $T\approx T_0$, with the tunnelling rate increasing sharply around this temperature.  A sudden increase of $\Gamma$ results in larger values of $\beta_*$, as per Eq.~\eqref{eq:beta}. For a plot of the tunnelling rate as a function of the temperature in this case ($m_0^2>0$), see App.~A in Ref.~\cite{Brummer:2025inh}.

After the phase transition completes, the Universe is reheated to a temperature $T_{\rm reh}$ such that $g_{\star,{\rm reh}}T_{\rm reh}^4 = g_{\star,*}T_*^4  + g_{\star,{\rm eq}}T_{\rm eq}^4$, upon assuming an instantaneous transition from vacuum domination to radiation domination (the two summands correspond to the contribution of radiation and vacuum to the energy density of the Universe just above $T_*$). In general, if the scalar field does not decay efficiently after the transition completion, the Universe may undergo an intermediate period of matter domination before reheating. However, we find that the scalar field promptly decays after the phase transition in the region of parameter space we are interested in, $\Gamma_{\rho\rightarrow H^\dagger H} \gg H(T_*)$, thus justifying the assumption of instantaneous reheating. 

Bubble collisions and plasma motion at the end of a sFOPT  are expected to generate a large signal of GWs, peaked at a characteristic frequency $\hat f \propto T_{\rm reh}\beta_*/H_{\rm reh}$, and with a peak amplitude $\Omega_{\rm GW}(\hat f) h^2 \propto (H_{\rm reh}/\beta_*)^{2}$. 
Details on the computation of the GW signal are given in App.~\ref{app:GWs}. In Fig.~\ref{fig:MonoFermsFOPT} we show the monopole relic density produced by a supercooled transition, together with the projected sensitivities of the Einstein Telescope (ET) \cite{ET:2019dnz} and the Cosmic Explorer (CE) \cite{Reitze:2019iox}, obtained assuming ten years of data-taking and a signal-to-noise ratio of five \cite{NANOGrav:2023hvm}. 

We find that, for values of $\eta$ associated to monopole DM, the characteristic frequency of GWs is typically in the  $(10^4 - 10^5) \;{\rm Hz}$ region, which is too large to be probed by upcoming interferometers like the CE and the ET, whose sensitivity peaks at $f\sim 10$ Hz. However, if $\hat f \sim 10^4\;{\rm Hz}$ and the amplitude of the GW spectrum is large enough, its low-frequency tail may be within reach, as shown in Fig.~\ref{fig:GWs}.\footnote{
For an earlier study of the correlation between monopole abundance and GW production, see Ref.~\cite{Yang:2022quy}. There, the authors focus on wFOPTs, concluding that, if one requires the monopoles produced by the phase transition to be the DM, the associated signal of GWs is many orders of magnitude below the sensitivity of the ET and the CE.} This happens, in particular, close to the ``No completion'' line in the left-hand panel in Fig.~\ref{fig:MonoFermsFOPT}. There, $\beta_*/ H_{\rm reh} \sim \mathcal{O}(0.1)$, while this ratio takes larger values away from the line. At smaller $\eta$, $T_{\rm reh}$ is correspondingly reduced and $\hat f$ is shifted to lower values even for large $\beta_*/H_{\rm reh}$; there, however, monopole abundance is suppressed. We find that, in the whole parameter space shown in the left panel, sound waves represents the dominant GW source. In the right-hand panel of Fig.~\ref{fig:MonoFermsFOPT}, $\beta_*/ H_{\rm reh}$ is typically large, due to $m_0^2 >0$, thus resulting into higher frequencies and smaller amplitudes. There, bubble collisions dominate the GW production for $g_0 \lesssim 0.6$. In this scenario, the monopole DM line is beyond the reach of future interferometers.

\subsection{Monopole annihilation}\label{sec:MonoAnn}
%%%%%%%%%%%%%%%%%%%%%%%%%%%%%%%%%%%%%%%%%%%%%%%%%%%%%%%%

After production, the monopole number density may still be reduced by monopole-antimonopole annihilations \cite{Zeldovich:1978wj,Preskill:1979zi} (see also Ref.~\cite{Vilenkin:2000jqa} for a review). Monopoles can dissipate their energy by moving through a plasma of relativistic charged particles, in our case the $e'$ fermions.\footnote{Energy loss from bremsstrahlung can be shown to be negligible for sufficiently heavy monopoles \cite{Vilenkin:2000jqa}.} This allows the formation of monopole-antimonopole bound states, which will eventually annihilate in less than a Hubble time.

At a temperature $T_{\rm capt} \approx \eta g^3/ (16\pi \mathcal{B}^2)$, the typical distance at which a bound state is formed becomes smaller than the monopole mean free path in the plasma. When this happens, the annihilation process freezes out. For a plasma of Dirac fermions with charge $1/2$, $\mathcal{B} = \zeta(3)/\pi^2$. Alternatively, annihilation stops being efficient soon after the dark electrons become non-relativistic. The temperature setting the final monopole relic density after annihilation, $T_a$, is therefore given by
\be
    T_a = \max \left[T_{\rm capt},\, m_{e'}\right].
\ee

In the case that it is monopole-antimonopole annihilation which determines the final monopole yield, this is given by \cite{Preskill:1979zi}
\begin{equation}\label{eq:YAnn}
    Y_M = \frac{\mathcal{B} g^2}{16\pi}\gamma_\star^{-1/2}\frac{T_a}{\MP}.
\end{equation}

Such annihilations may efficiently reduce the abundance of monopoles produced at the phase transition. This is the case in the scenarios illustrated in Fig.~\ref{fig:SOPTFermi} (SOPT) and Fig.~\ref{fig:MonoFermwFOPT} (wFOPT): for sufficiently small $g$, the monopole abundance follows Eq.~\eqref{eq:YAnn} independently from the initial production mechanism.  On the other hand, the monopole number density produced during a sFOPT (illustrated in Fig.~\ref{fig:MonoFermsFOPT}) is small enough for annihilations not to be effective after the phase transition.

A very recent reanalysis of the monopole annihilation mechanism \cite{Liveoak:2026muy} obtains a slightly larger final abundance, which implies a slightly larger window of parameters to realise monopole DM.

\section{Phenomenological constraints}
\label{sec:CosmoConst}

In this section we discuss some experimental signatures that one might expect to constrain our model, other than those coming from dark radiation and GWs, already discussed above. Unfortunately, we will see that present-day experiments are unable, by a large margin, to probe the parameter region where the monopoles constitute all of the DM. In short, monopoles are far too heavy for conventional DM searches, and the light dark-sector particles are too weakly coupled to the SM to be probed by collider searches.

\subsection{Higgs decay into invisible}\label{sec:InvDec}

The branching ratio of the SM Higgs decaying into invisible particles is constrained by collider searches to be ${\rm BR}(h \rightarrow {\rm inv}) < 0.107$ at the 95\% CL \cite{ATLAS:2023tkt}.  The total decay width of the Higgs is $\Gamma_{h,{\rm tot}} = 3.7\,{\rm MeV}$ \cite{ParticleDataGroup:2024cfk}, leading to $\Gamma(h \rightarrow {\rm inv})< 0.396\,{\rm MeV}$. 

An on-shell Higgs boson can decay into a pair of dark, massless photons, via mixing with the dark sector radial mode. The rate for such a process, given in Eq.~\eqref{eq:hgammagamma}, is however loop suppressed. A tree-level decay into dark electrons is also possible, cfr.~Eq.~\eqref{eq:hee}, provided that $m_{e'}<m_h/2$. In the region of parameters of interest to us, the dark electrons are light, $ 50\;{\rm MeV}\lesssim m_{e'} \lesssim 50\;{\rm GeV}$, and the process $h \to e' \bar{e}'$ represents the dominant decay channel of the SM Higgs into dark sector particles. 

The region excluded by the upper bound on the Higgs invisible width is shown as a black-dotted region in Fig.~\ref{fig:PhenoConstr}, for the cases of a SOPT and a wFOPT.

\subsection{Self-interacting dark matter}\label{sec:SIDM}

Dark monopoles interact via a long-range, Coulomb-like magnetic force and are therefore an example of self-interacting DM. 
DM elastic scattering may help solving some long-standing puzzles of the collisionless DM paradigm, such as the “cusp-core" problem
(see Refs.~\cite{Tulin:2017ara,Adhikari:2022sbh} and references therein). However, astrophysical observations on both small and large scales, ranging from Bullet Cluster observations to ellipticity of galaxies, pose upper bounds on the cross section of DM self interaction.

In the classical and non-relativistic limit, 
the monopole-antimonopole elastic scattering is described by the transfer cross section
\cite{Tulin:2013teo},
\be\label{eq:sigmaT}
   \sigma_T = \frac{16 \pi \alpha_M^2}{m_M^2 v_M^4} \log\left(1+ \frac{m_M v_M^2}{2\alpha_M m_{\gamma'}}\right).
\ee
where $\alpha_M = 4\pi/g^2$ is the magnetic analog of the fine-structure constant,  $v_M$ is the local velocity of monopoles at the relevant scales, 
$m_{\gamma'}^2=16\pi \alpha_M \rho_M/(v_M m_M)^2$ is the Debye mass of the dark photon \cite{Kahlhoefer:2013dca}, with $\rho_M$ the local monopole density. Notice that, since $\alpha_M/m_M\approx r_M$, this cross section is of the order of a geometric cross section.

\begin{figure}[tb!]
\begin{center}
\includegraphics[width=.48\textwidth]{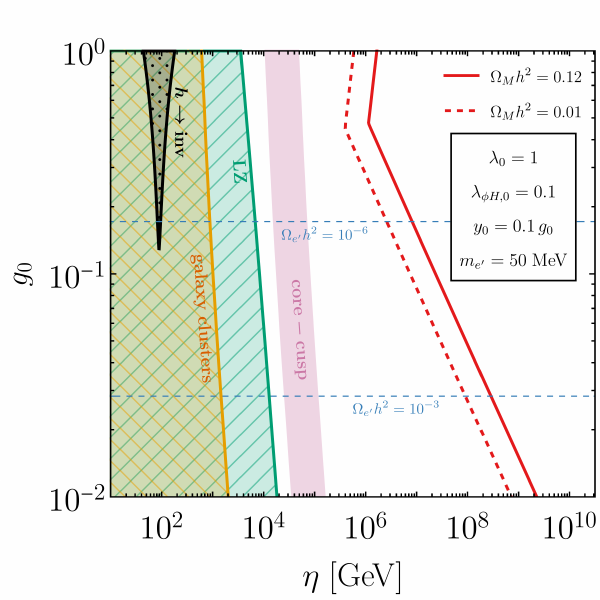}\quad\includegraphics[width=.48\textwidth]{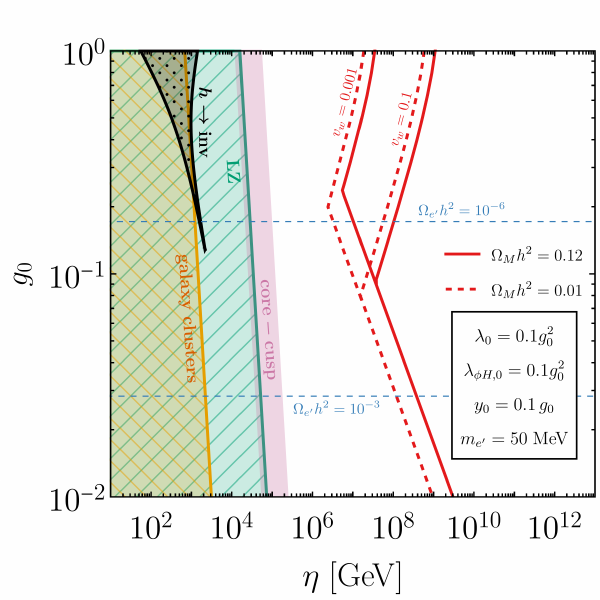}
    \caption{
    Phenomenological constraints on monopole DM in the scenario of a SOPT (\textit{left panel}) and of a wFOPT (\textit{right panel}). The black-dotted region is excluded by bounds on the Higgs invisible decay width (Sec.~\ref{sec:InvDec}). The orange, hatched region is excluded by monopole self-interactions that are too large to be consistent with DM profile observations at galaxy-cluster scales (Sec.~\ref{sec:SIDM}); in the pink shaded region, monopole self-interaction may solve the core-cusp problem. The green, hatched region is excluded by the LUX-ZEPLIN direct detection experiment (Sec.~\ref{sec:DirectDet}).  As usual, red (blue) lines show isocontours of the relic abundance of monopoles (dark electrons).
    }
\label{fig:PhenoConstr}  
\end{center}
\end{figure}

We find that the tightest bound on monopole self-interaction comes from galaxy-cluster scales. Ref.~\cite{Andrade:2020lqq} uses strong-lensing observations of  clusters with an average velocity of $v_M \simeq 1.5\times 10^3 {\;\rm km/s}$ to infer the DM profile and to set $\sigma_T/m_M < 0.13\;{\rm cm^2/g}$ at the two sigma level. Ref.~\cite{Sagunski:2020spe} finds $\sigma_T/m_M < 0.35\;{\rm cm^2/g}$ at similar scales. 
Constraints from galaxy clusters are shown in orange  in Fig.~\ref{fig:PhenoConstr}. 

 On the other hand, a self-interaction cross section $ 0.5 \,{\rm cm}^2/{\rm g} \lesssim \sigma_T/m_M \lesssim 50\,{\rm cm}^2/{\rm g}$, at velocities $v_M\simeq 40\,{\rm km}/{\rm s}$, could solve the “cusp-core" problem by alleviating the tension between observations and numerical simulations \cite{Elbert:2014bma}. The region of parameters for which $\sigma_T/m_M = \left(0.5 - 50\right) \,{\rm cm}^2/{\rm g}$  is shown in pink in Fig.~\ref{fig:PhenoConstr}.

\subsection{Direct detection}\label{sec:DirectDet}
At energies below $r_M^{-1}$, the monopole can be treated as a point-like object. In an effective field theory containing a dynamical Higgs, at energies above $m_h$, the monopole trilinear coupling to the Higgs boson reads \cite{Bai:2020ttp}
\be
    \mu_{M\bar M h}=\left.\frac{\partial m_M^2(h)}{\partial h}\right|_{h=v} = \left(\frac{4\pi k}{g}\right)^2\frac{\lambda_{\phi H}}{\lambda}v \ .
\ee
The elastic scattering of a dark monopole off a nucleon $N$ is mediated by a Higgs boson. In the limit of small momentum exchange, non-relativistic monopoles, and $m_M\gg m_N$, the cross section for such a process is 
\be
    \sigma_{MN\to MN}\approx \frac{y_{h N N}^2}{4\pi} \frac{\mu_{M\bar M h}^2 m_N^2}{m_h^4 m_M^2}\approx \frac{4\pi k^2 y_{h N N}^2 \lambda_{\phi H}^2}{\lambda^2} \frac{m_N^2 v^2}{m_h^4}  r_M^2 \ ,
\ee
where $m_N \simeq 0.94\;{\rm GeV}$ is the nucleon mass, and $y_{hNN} \equiv f_N m_N/v$ is the effective coupling of the Higgs to nucleons, with $f_N \simeq 0.3$ \cite{Bishara:2017pfq}.

DM direct detection experiments set bounds on the DM-nucleon elastic cross section up to very heavy DM candidates, as it is the case for dark monopoles. The leading experimental constraint is the one reported by the LUX-ZEPLIN (LZ) collaboration \cite{LZ:2024zvo,LZ:2024psa}, which we show in green in Fig.~\ref{fig:PhenoConstr}. In the scenario of a SOPT, our computation of the monopole abundance suggests a lower bound $m_M\gtrsim 10^8$ GeV, in order for monopoles to account for the totality of DM. For this mass, the LZ constraint is $\sigma_{MN \to MN}^{\rm LZ} \lesssim 3\times 10^{-42}\;{\rm cm}^2$ at the 90\% CL. For comparison, a monopole DM population produced by a SOPT with $m_M = 10^8\;{\rm GeV}$ gives $\sigma_{MN \to MN} \simeq 10^{-49}\;{\rm cm}^2$, where we have chosen $\lambda_0 =1$ and $\lambda_{\phi H, 0} = 0.1$ to make contact with Fig.~\ref{fig:SOPTFermi}. Larger values of $m_M$ are associated to even smaller cross sections. In the case of a wFOPT, we find that monopole DM cannot be lighter than $m_M \simeq 3 \times 10^8\;{\rm GeV}$, where the LZ bound reads $\sigma_{MN\to MN}^{\rm LZ} \simeq 9\times 10^{-42}\;{\rm GeV}^2$ at the 90\% CL. Fixing $\lambda = \lambda_{\phi H} = 0.1 g^2$, as we did in Fig.~\ref{fig:MonoFermwFOPT}, the monopole-nucleon scattering cross section is smaller than $\sigma_{MN - MN}\simeq 10^{-48}\;{\rm cm}^2$. Finally, monopole DM from a sFOPT is typically much heavier, thus resulting in even looser direct-detection constraints. We conclude that, even in the most optimistic scenario, the cross section for monopole-nucleon elastic scattering is about 7 orders of magnitudes below the current experimental reach.

\subsection{Indirect detection}

DM particles can annihilate into SM states in dense environments such as the galactic center or the inner galactic halo. The predicted flux of SM stable particles produced by DM annihilation, after subtraction of astrophysical backgrounds, can be tested against experimental constraints.

Monopole-antimonopole annihilation into SM particles is a non-perturbative process which we will not aim to model in detail; we will rather provide arguments that the resulting signal is far too small to be of phenomenological interest. Following Ref.~\cite{Drukier:1981fq}, the 't Hooft-Polyakov monopole can be understood as a coherent state made of a large number $N={\cal O}(4\pi/g^2)$ of quanta of $\phi$ and the dark gauge field. Consider now monopole pair production, $XX'\into M\overline M$, where $X$ and $X'$ are SM states. This necessitates first producing $2N$ dark-sector quanta and then having them collapse into a monopole-antimonopole pair. The latter process cannot be treated in perturbation theory. But the former can, since the dark-sector fields are weakly coupled to each other and to the SM. The amplitude for producing $2N$ dark-sector quanta is exponentially suppressed, scaling as $g^{2N}$. Ref.~\cite{Drukier:1981fq} concludes that pair production of 't Hooft-Polyakov monopoles at high-energy colliders is not a realistic possibility, even if it were possible kinematically.

Consider now the inverse process $M\overline M\into X X'$; in the most relevant case, $X=X'=h$ is the SM Higgs boson, since the Higgs portal coupling is the only one connecting the SM with the dark sector. By an analogous argument, the amplitude is exponentially suppressed. Note that this suppression applies only to the exclusive $hh$ final state, while the amplitude for the inclusive process $M\overline M\into hh\,+\,({\cal O}(2N)$ dark-sector particles) should only be suppressed by 
$\lambda_{\phi H}$ and possibly some non-perturbative form factor (estimating the overall $M\bar M$ annihilation probability).\footnote{For illustration, compare this with a $p\bar p$ collision at high energies such that $\alpha_s\ll 1$; the amplitude for the exclusive process $p\bar p\into e^+ e^-$ is tiny, while the amplitude for $p\bar p\into e^+ e^-\,+\,$(hadrons) is suppressed only by $\alpha_{EM}$.}  Nevertheless, the energy going into SM particles from such a process is $\ll 2 m_M$,  and in addition the monopole number density is small  for large $m_M$, so the resulting flux of SM particles is reduced and their energies are much lower than the DM mass.

More quantitatively, indirect detection experiments targeting very heavy DM are sensitive to cross-sections $\langle\sigma v\rangle\gtrsim 10^{-22}$ cm$^3/$s for DM masses $\gtrsim 10^8$ GeV (see e.g.~\cite{Boehm:2025qro} for LHAASO bounds with a $hh$ final state). This alone is four orders of magnitude above a (conservatively estimated) geometric total $M\overline M$ annihilation cross-section for $m_M\approx 10^{8}$ GeV and $g\approx 10^{-2}$. 
With the additional suppression according to the above discussion, we clearly cannot expect a detectable signal in indirect detection from monopole DM.

\section{Discussion}

We have shown that the DM of the Universe can be made of dark-sector 't Hooft-Polyakov monopoles, in a non-minimal ultraviolet-complete model. This was achieved by adding light fermions to the minimal dark sector. 

The question arises to what extent the model parameters need to be fine-tuned for our mechanism to work. In particular, to obtain the observed relic density, dark-sector spontaneous symmetry breaking has to occur at a scale $\eta$ much larger than the electroweak scale. We have nothing new to add on the usual scalar-field hierarchy problem; this separation of scales is radiatively unstable in the presence of a sizeable Higgs portal coupling $\lambda_{\phi H}$ and in the absence of a mechanism to shield the electroweak scale from quantum corrections. Secondly, bounds on dark radiation set strong constraints on the dark fermion parameters. In particular, the dark muon is required to be  much heavier than the dark electron, and hence a partial cancellation between the dark Higgs-dependent ($y\eta$) and the field-independent ($m_f$) contributions to the dark fermion masses. This implies a tuning of the order of $m_{e'}/(y\eta)$, which can be mitigated by taking the (technically natural) limit of a very small Yukawa coupling $y$.\footnote{
In the previous sections we did not try, for simplicity, to minimise the value of the Yukawa $y$, and we just tuned parameters correspondingly. We remark that the analysis of the monopole abundance remains qualitatively the same in the limit of very small $y$.}
Yet the dark muon must remain sufficiently heavy to decay before the dark and visible sectors decouple at the temperature $T_{\rm dec}$. 
Given that (i) $\mu'$ decays at a temperature around $m_{\mu'}\sim y\eta$, (ii) $T_{\rm dec}\sim 10^5\,{\rm GeV}$ for typical values of 
$\eta$, $g^2$, $\lambda$ and $\lambda_{\phi H}$, and (iii) $m_{e'} \lesssim 10^2\,{\rm GeV}$, we are left with a minimal fine-tuning in the fermion sector of the order of $m_{e'}/T_{\rm dec}\lesssim 10^{-3}$.  This is the price to suppress electrically-charged relics and dark radiation, in the region of parameters where monopoles constitute the DM.

It is remarkable that our model lives close to the present-day exclusion limits from dark radiation. If the BBN bounds should consolidate and improve in the near future, the model must be considered ruled out, with the exception of a small window of parameters in the case of a wFOPT, see right panel of Fig.~\ref{fig:MonoFermwFOPT}. Moreover, the Simons Observatory is expected to considerably improve the CMB constraints in the next 10 years, with an expected one-sigma sensitivity on $N_{\rm eff}$ of $\sigma_{N_{\rm eff}} \sim 0.045$ \cite{SimonsObservatory:2025wwn}, which would also rule out the model. If, conversely, evidence for $\Delta N_{\rm eff}$ exceeding the SM prediction should be found, this would motivate 
a more detailed study of the model phenomenology in other channels. 
We showed that the (current) sensitivities of conventional DM searches are orders of magnitude away from the parameter space regions relevant for monopole DM. Therefore, the best prospects to further constrain the model might be with GW searches, probing the dark phase transition if it is strongly first-order. Several proposals have been put forward to probe the high-frequency region of GWs relevant to monopole DM \cite{Aggarwal:2020olq}. Their sensitivities, however, are too small to probe signals of cosmological origin, and a sizeable improvement seems unfeasible with current technologies \cite{TitoDAgnolo:2024res}.

While the numerical studies we undertook are, of course, specific to the model we studied, let us stress that our results should apply much more generally on a qualitative level. Any dark-sector model with stable magnetic monopoles from spontaneous symmetry breaking will feature a massless dark photon (contributing to dark radiation) and charged massive gauge bosons (potentially contributing to DM). If the thermal relic density of the latter is to be subdominant, they need to be diluted somehow or be unstable, ultimately decaying into the lightest electrically-charged dark sector particle. If this particle is light, it will have an additional impact on $\Delta N_{\rm eff}$.
One might ask whether there are non-minimal models which are less constrained than ours. But such models are difficult to envisage (to say the least), if they are to contain other light dark-sector particles. One light dark fermion along with the massless dark gauge boson is already in tension with the dark radiation constraints from BBN. A light charged scalar might be present instead but would hardly do better, resulting in $\Delta N_{\rm eff}\approx 0.135$ in the large $T_{\rm dec}$ limit. Any additional light particles heating up the dark sector below $T_{\rm dec}$ would be in tension already with the more conservative CMB bound. 

It is of course  possible to relax some of our basic assumptions, at the price of introducing different free parameters, and thus  losing some predictivity: the dark sector might never have been in equilibrium with the SM; the monopoles might be created by some mechanism other than a thermal phase transition; the dark-sector phase transition might interfere in a non-trivial way with the electroweak one; or, the monopoles might have a microscopic origin which cannot be described by ordinary (weakly coupled, four-dimensional) field theory. A quantitative analysis of complete models along these lines has yet to be developed.

\subsection*{Acknowledgments}
%%%%%%%%%%%%%%%%%%%%%%%%%%%%%%%%%%%

The authors thank M.~Cirelli, M.~Escudero Abenza, J.~Froustey, Y.~Gouttenoire, S.~Manconi, S.~Rosauro-Alcaraz, A.~Stanzione, and P.~Schicho for useful discussions. 
MF has received support from the European Union Horizon 2020 research
and innovation program under the Marie Sklodowska-Curie grant agreement No 101086085–ASYMMETRY.

\appendix 
\counterwithin*{equation}{section}

\section{Running couplings}\label{app:betaFunc}
\renewcommand\theequation{\thesection\arabic{equation}}

Here we provide the one-loop $\beta$ functions governing the evolution of the couplings with the renormalisation scale, as well as the one-loop anomalous dimension $\gamma_\phi$ of the dark scalar field. We have checked our results with the \texttt{PyR@TE} \cite{Sartore:2020gou} and \texttt{RGBeta} \cite{Thomsen:2021ncy} public codes. Defining
\be
 \mu \frac{d c_i}{d \mu} = \frac{\beta_{c_i}}{16\pi^2}\ , \qquad \mu \frac{d \phi}{d\mu} = -\frac{\gamma_\phi}{16\pi^2} \phi\ , 
\ee
with $c_i=\left(\lambda,\,g,\,y,\, m_f,\,m^2,\lambda_{\phi H}\right)$, we find in the unbroken phase of the dark $\SU{2}$
\begin{align}
    \label{eq:betag}
    \beta_g&= -\frac{19}{3} g^3 =\left(-22\rvert_g+1\rvert_s+2\rvert_f\right)\frac{g^3}{3}\ ,\\
    \beta_\lambda&=22\lambda^2-24\lambda g^2+12g^4+16 \lambda y^2  -16y ^4 +2\lambda_{\phi H}^2\ ,\\
    \beta_y&=\left(5y^2-\frac{9}{2}g^2\right)y\ ,\\
    \beta_{m_f}&=\left(9y^2-\dfrac 92 g^2\right)m_f\ ,\\
    \beta_{m^2}&\simeq(10\lambda-12g^2+8y^2)m^2+48 y^2m_f^2\ ,\\
    \beta_{\lambda_{\phi H}} & \simeq \lambda_{\phi H}\left(10\lambda-12g^2+8y^2+4\lambda_{\phi H}
    -\frac{9}{2}g_W^2-\frac{3}{2}g_Y^2+6y_t^2 \right)\ ,\\
    \gamma_{\phi} & =-6g^2 + 4y^2\ ,
\end{align}
where we have indicated for convenience the separate contributions of gauge, scalar and fermion loops to $\beta_g$, in $\beta_{m^2}$ we have included only the dark-sector contributions (neglecting a term $\sim \lambda_{\phi H}\mu_H^2$ involving the small SM Higgs mass parameter), and  in $\beta_{\lambda_{\phi H}}$ we have included only the leading SM contributions (neglecting all but the top quark Yukawa couplings, as well as the Higgs self-coupling).
The wavefunction renormalisation of the scalar field, $Z_\phi$, defined by $\phi(\mu) = Z_{\phi}^{-1/2}(\mu) \phi_0$, is related to the anomalous dimension through
\be\label{eq:Zphi}
Z_\phi(\mu) = \exp\left(\frac{2}{16\pi^2} \int_{\mu_0}^\mu \frac{d\mu'}{\mu'}\gamma_{\phi}\right).
\ee

At the scale $\mu_0 =m_{W',0}=g_0 \eta$, the dark $\SU{2}$ gauge symmetry is spontaneously broken to $\U{1}$, and the $W'$ is integrated out. Properly speaking, the scalar $\rho$ should be integrated out at its own mass scale, $m_\rho$, but for simplicity we decouple it at $\mu_0$, the difference being numerically negligible.
The two Dirac fermions have masses $m_{\mu'}=m_f+y\eta$ and $m_{e'}=m_f-y\eta$, and they
should contribute to the running only above their respective mass thresholds; since we always assume $\mu'$ to be heavy, we also decouple it at $\mu_0$. At low scales, the theory becomes dark QED with a single Dirac fermion $e'$ and a dark photon. The dark electron has an effective dimension-5 interaction with the SM Higgs, whose Wilson coefficient is obtained by matching with the ultraviolet 
theory, 
$c_{e'e'HH,0}=y_0\lambda_{\phi H,0} \eta /m_{\rho,0}^2$.

The $\beta$ functions for the dark QED coupling (still called $g$ for simplicity), the $e'$ mass, and the $e'-H$ portal are given by
\begin{align}
    \beta_g&=\frac{1}{3}g^3\ ,\\
    \beta_{m_{e'}}&=-\frac{3}{2}g^2m_{e'}\ ,\\
    \beta_{c_{e'e'HH}}&\simeq-\frac{3}{2}c_{e'e'HH}\left(g^2+3g_W^2+g_Y^2-4y_t^2\right)\ .
\end{align}

\begin{figure}[tb!]
\begin{center}
\includegraphics[width=.48\textwidth]{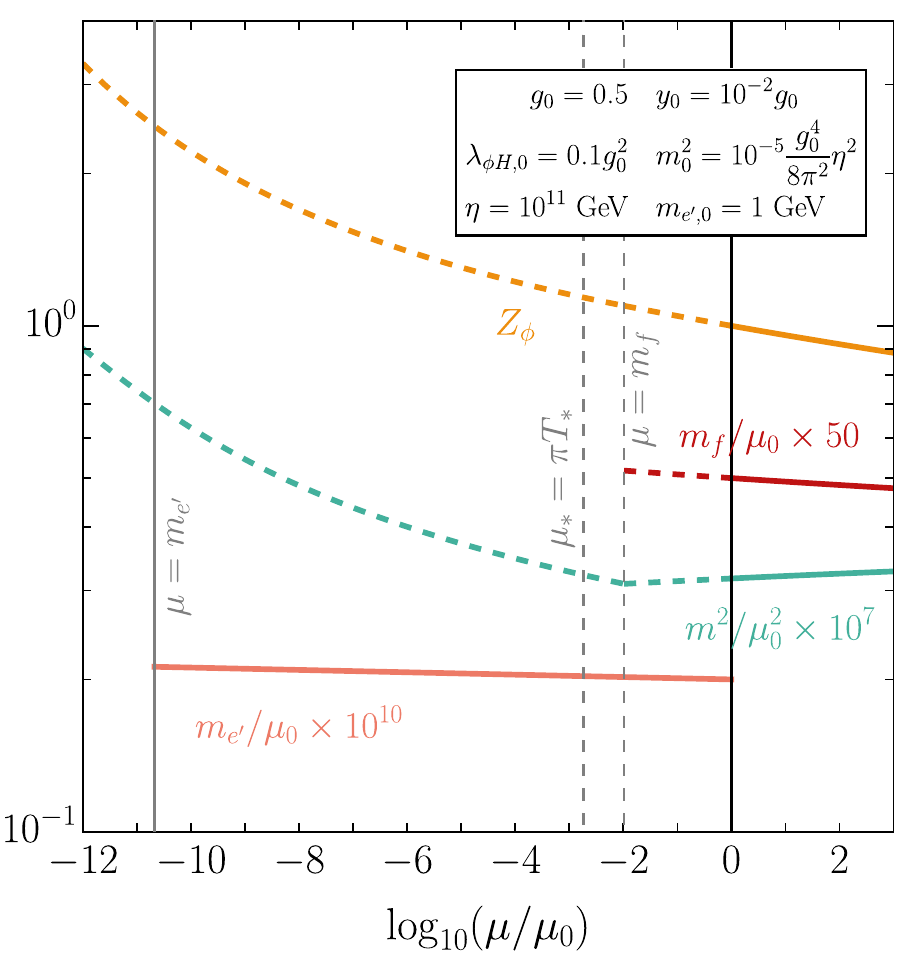}\quad \includegraphics[width=.48\textwidth]{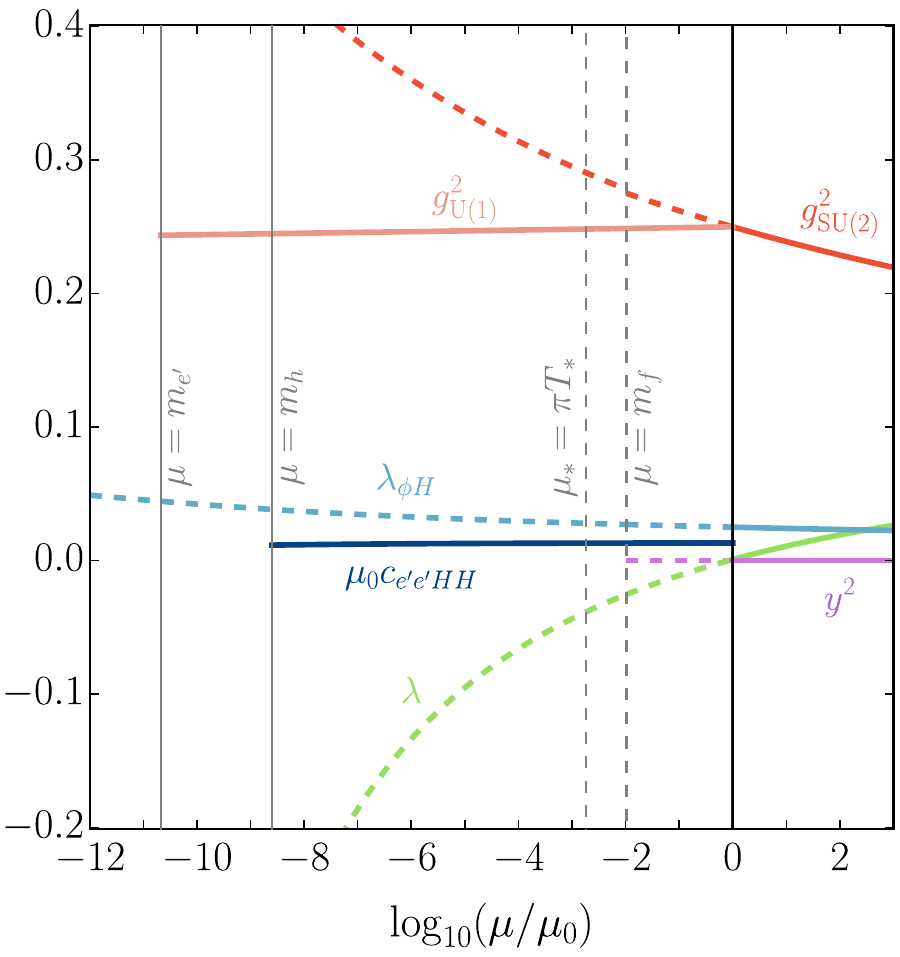}
    \caption{Running couplings as a function of the renormalisation scale $\mu$, for a choice of parameters corresponding to a point in the right-hand panel of Fig.~\ref{fig:MonoFermsFOPT}.  The values of the parameters in the legend are defined at the reference scale $\mu_0=g_0\eta$, and refer to both panels; the Coleman-Weinberg relation is assumed to fix $\lambda_0$. The unbroken theory couplings are represented by solid (dashed) lines above (below) $\mu_0$: dashed lines are relevant when the Universe is stuck in the false vacuum down to small scales (in this case fermions are decoupled at $m_f$). The broken theory couplings are described by solid lines below $\mu_0$ (in this case $W'$, $\rho$ and $\mu'$ are all decoupled at $\mu_0$ for simplicity). The \textit{left panel} illustrates the running of the mass scales and of the scalar wavefunction, while the \textit{right panel} shows the running of the dimensionless couplings and of the effective coupling between dark electrons and Higgs doublets. 
    The dashed vertical line at $\mu=\mu_*$ indicates the scale at which the transition to the true vacuum  completes, for this choice of parameters. 
    }
\label{fig:betafunc}  
\end{center}
\end{figure}

The above renormalisation group equations above are solved numerically. The solution is shown in Fig.~\ref{fig:betafunc} for a benchmark set of parameters, corresponding to the ones used in Fig.~\ref{fig:MonoFermsFOPT}. We have fixed $\lambda$ to its Coleman-Weinberg value at the scale $\mu_0$, according to Eq.~\ref{eq:CWrelation}: this is our benchmark for a supercooled phase transition, which is the scenario where the running of the couplings matters the most.

Note that, in the case of a supercooled phase transition, the Universe remains stuck in the false vacuum at temperatures well below $\mu_0$, until it eventually tunnels at some small $T_*$. In the false vacuum, $\SU{2}$ remains unbroken, with $\eta=0$, massless gauge bosons, very light 
dark scalars (since $m^2$ must be small to allow for supercooling), and two degenerate Dirac fermions with mass $m_f$. The dashed lines in Fig.~\ref{fig:betafunc} illustrate the running of the couplings in the false vacuum below the scale $\mu_0$.

\section{Thermalisation between dark and visible sectors}
\label{app:ThermHist}
\renewcommand\theequation{\thesection\arabic{equation}}

Given a process of the kind $a\,b\,\cdots \leftrightarrow \alpha\,\beta\,\cdots$, where letters from the beginning of the Latin (Greek) alphabet denote particles in the initial (final) state, the Boltzmann equation governing the comoving number density of the $a$-th particle species is
\be
    s H z_a \frac{d Y_a}{d z_a} = N_a\, \gamma_{\rm eq}(a\,b\,\cdots \leftrightarrow \alpha\,\beta\,\cdots)\left(\frac{Y_a}{Y_a^{\rm eq}}\frac{Y_b}{Y_b^{\rm eq}}\cdots - 1\right),
\ee
with $N_a$ the number of $a$ particles created minus the number of $a$ particles destroyed in the process, and $z_a \equiv m_a/T$. In the above expression, we have assumed the final particles to be in thermal equilibrium. The \textit{reaction density} $\gamma_{\rm eq} (a\,b\,\cdots \leftrightarrow \alpha\,\beta\,\cdots)$ represents the rate of the process $a\,b\,\cdots \leftrightarrow \alpha\,\beta\,\cdots$ per unit volume, and it is defined as
\be\label{eq:gamma}
    \gamma_{\rm eq}(a\,b\,\cdots \leftrightarrow \alpha\,\beta\,\cdots) \equiv \frac{1}{n_i!n_f!} \int \prod_i d^3\bar p_i\,f_i^{\rm eq}  \int \prod_f d^3\bar p_f(2\pi)^4\delta^4\left(\sum_i p_i  - \sum _f p_f \right) \lvert\mathcal{M}\rvert^2,
\ee
where $d^3\bar p\equiv d^3p /\left[2 E(2\pi)^3\right]$, and the indices $i = a,b,\cdots$ and $f=\alpha,\beta,\cdots$ are for particles in the initial and final state, respectively. Here $n_i$ ($n_f$) is the number of identical initial (final) particles. The quantity $\lvert \mathcal{M} \rvert^2$ is the transition amplitude summed over all the initial and final indices (spin, gauge, etc.).

In the following, we will focus on two-body decays $1 \rightarrow 2\,3$, and on two-to-two scatterings  $1\,2\rightarrow 3\,4$. For the former, Eq.~\eqref{eq:gamma} reduces to
\be
    \gamma\left(1 \leftrightarrow 2\,3\right) = g_1 \frac{m_1^2\,T}{2\pi^2} K_1\left(\frac{m_1}{T}\right) \Gamma(1\rightarrow 2\,3) \,,
\ee
where $K_1$ is the Bessel function of the first kind, $g_1$ is the number of internal degrees of freedom of particle 1, and $\Gamma$ is the decay rate computed by summing over final states and averaging over the initial state, and including a $1/2!$ factor if particles 2 and 3 are identical.  For a two-to-two scattering, Eq.~\eqref{eq:gamma} gives \cite{Gondolo:1990dk}
\begin{equation}
    \gamma \left(1\,2 \leftrightarrow 3\,4 \right) = \frac{T}{64 \pi^4} \int_{s_{\text{min}}}^{+\infty}ds \sqrt{s}\,\hat{\sigma}(s) K_1 \left(\frac{\sqrt{s}}{T} \right),
    \label{eq:gammascat}
\end{equation}
with $s_{\text{min}}=\max \left[(m_1+m_2)^2,\, (m_3+m_4)^2 \right]$ and 
\begin{equation}
\hat{\sigma}\left(1\,2\leftrightarrow 3\,4 \right)= \frac{g_1g_2}{n_i!}\frac{2}{s}\left[(s-m_1^2-m_2^2)^2-4m_1^2m_ 2^2 \right] \sigma\left(1\,2\rightarrow 3\,4 \right)\,.
\end{equation}
Here, $\sigma$ is the usual cross-section, averaged over the initial state and summed over the final state, and containing the proper symmetry factors for identical final-state particles.

\begin{figure}[tb!]
\begin{center}
\includegraphics[width=0.55\textwidth]{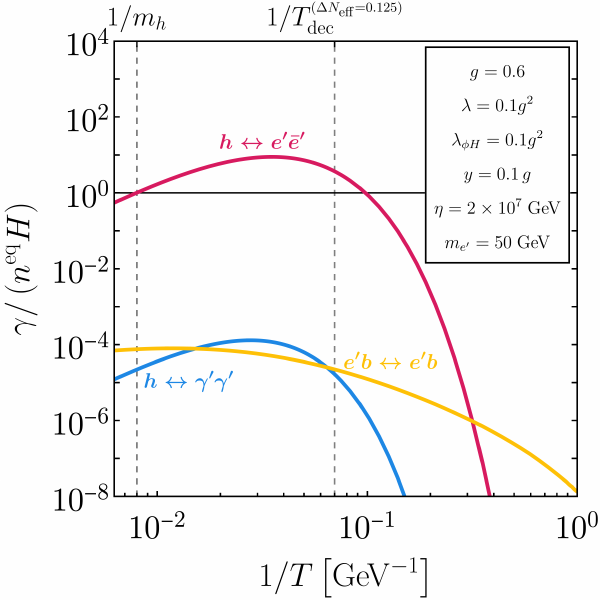}
    \caption{
    Interaction rate for various dark-visible sector reactions, normalised to the dark-particle equilibrium number density times Hubble constant, as a function of the temperature. 
    With this normalisation, the dark particle is kept in equilibrium when the rate is larger than one (horizontal black line).
    The dashed, vertical line at $T = T_{\rm dec}^{\scriptscriptstyle (\Delta N_{\rm eff} = 0.125)}$ corresponds to the largest decoupling temperature allowed by BBN+CMB constraints on dark radiation. For the chosen values of the model parameters, $T_{\rm dec} \simeq 10\;{\rm GeV}$ and  $\Delta N_{\rm eff} \simeq 0.097$, and monopoles comprise a sizeable fraction of DM, see right panel of Fig.~\ref{fig:MonoFermwFOPT}.}
\label{fig:ReacDens}  
\end{center}
\end{figure}

The species $a$ is kept in equilibrium by a given reaction  as long as 
\be
    \frac{N_a\,\gamma_{\rm eq}(a\,b\,\cdots\leftrightarrow \alpha\,\beta\,\cdots)}{n_a^{\rm eq} H} >1 \,.
\ee
In the rest of this appendix, we will study various reactions between SM and dark sector particles, to determine which process is the most relevant to achieve and sustain thermalisation between the two sectors. We focus on temperatures below the electroweak phase transition, as our main concern is the decoupling of dark electrons, which are lighter than the electroweak scale.
Our findings are summarised in Fig.~\ref{fig:ReacDens}: 
dark electrons can be kept in equilibrium with the SM down to temperatures a few times smaller than $m_{e'}$, as long as the portal is sufficiently large (both $y$ and $v/\eta$ not too small, see  Eq.~(\ref{eq:hee})). 

\subsection{$\boldsymbol{h\rightarrow e' \bar{e}'}$}

The dark sector scalar field $\rho$ decays into a pair of dark electrons via a Yukawa coupling. Due to the presence of the portal interaction, the $\rho$ mixes with the SM Higgs $h$, 
with mixing angle $\theta$ given by  
\be
    \tan 2\theta = \frac{2\lambda_{\phi H}\eta\,v}{m_\rho^2 - m_h^2} \,.
\label{theta}\ee
If $m_{e'}<m_h/2$, the SM Higgs can therefore decay into a an $e'\bar{e}'$ pair at tree level, with decay rate 
\be\label{eq:hee}
    \Gamma (h \rightarrow e'\bar e') = \sin^2\theta \, \frac{y^2}{8\pi} \, m_h \left(1- \frac{4 m_{e'}^2}{m_h^2}\right)^{3/2}\,.
\ee
Such decays (and inverse decays) are the dominant process to maintain the thermal contact between the SM and the dark electron, for temperatures $ T$ around $m_h$. Only when $T\ll m_h/10$ does the Higgs number density become so suppressed that scattering processes, such as
$ b \bar{b} \to e' \bar{e}'$ via off-shell Higgs exchange, come to dominate. However, in the region of parameters we are interested in, the dark electrons decouple before, and these scattering rates are always out of equilibrium.

\subsection{$\boldsymbol{h\rightarrow \gamma'\gamma'}$}
%%%%%%%%%%%%%%%%%%%%%%%%%%%%%%%%%%%%%%%%%%%%%%%%%%%%%%%%%%%%%%

Analogously to what happens in the SM, the dark sector scalar field $\rho$ decays into a pair of dark photons via a loop of dark fermions or a loop of dark gauge bosons. 
As the $\rho$ mixes with the SM Higgs $h$, the latter can decay into dark photons with width
\be\label{eq:hgammagamma}
    \Gamma\left(h \rightarrow \gamma'\gamma'\right) = \sin^2 \theta \frac{g^4}{4^6 \pi^5 }\frac{m_h^3}{\eta^2}\lvert F\left(m_h^2\right) \rvert^2\,,
\ee
where the loop function $F\left(p^2\right)$ reads \cite{Bizot:2015zaa} 
\be
    F\left(p^2\right) = A_1\left(\frac{p^2}{4m_{W'}^2}\right) + \frac{y\,\eta}{4 m_{e'}}A_{1/2}\left(\frac{p^2}{4m_{e'}^2}\right)+\frac{y\,\eta}{4 m_{\mu'}}A_{1/2}\left(\frac{p^2}{4m_{\mu'}^2}\right)\,,
\ee
with
\be
\begin{aligned}
    A_1(x) = - \frac{2x^2 + 3x + 3(2x-1) f(x)}{x^2}\,, \qquad A_{1/2}(x) = \frac{2\left[x+\left(x-1\right)f(x)\right]}{x^2}\,,
\end{aligned}
\ee
and
\be
    f(x)=\left\{\begin{array}{cl}
\arcsin ^2 \sqrt{x} & \text { for } x \leq  1, \\
-\frac{1}{4}\left(\log \frac{1+\sqrt{1-x^{-1}}}{1-\sqrt{1-x^{-1}}}-i \pi\right)^2 & \text { for } x>1 .
\end{array}\right.
\ee
As shown in Fig.~\ref{fig:ReacDens}, the corresponding reaction density is always much smaller than the one for $h\rightarrow e' \bar{e}'$. This means that dark photons do not thermalise directly with the SM, rather they are in thermal equilibrium with dark electrons by virtue of the dark $\U{1}$ interaction.
Scattering processes such as $b\bar b \to \gamma'\gamma'$ are subleading with respect to $h\to \gamma'\gamma'$, in analogy to the case of dark electrons, as explained in the previous subsection.

\subsection{$\boldsymbol{e'b\rightarrow e'b}$}
%%%%%%%%%%%%%%%%%%%%%%%%%%%%%%%%%%%%%%%%%%%%%%%%%

Efficient elastic scatterings are sufficient to maintain kinetic equilibrium between the two sectors, ensuring that they share a common temperature. Contrary to annihilations, such scatterings do not present a kinematic threshold, and they remain active even for temperatures below the masses of the particles involved. 

The relevant process which could maintain dark electrons in kinetic equilibrium at $T\ll m_h$ is the elastic scattering on bottom quarks, via a $t$-channel $h$ or $\rho$ exchange. The cross section reads
\begin{equation}
\sigma_{e'b\rightarrow e'b}(s) = \frac{y^2\sin^2(2\theta)}{4\pi}\frac{m_{e'}^2m_b^4}{v^2(s-m_{e'}^2-m_b^2)^2}\int_{t_{\rm min}}^0 dt \left(\frac{1}{t-m_\rho^2}-\frac{1}{t-m_h^2}\right)^2 \left(1-\frac{t}{4 m_{e'}^2}\right)\left(1-\frac{t}{4m_b^2}\right),
\end{equation}
with $t_{\rm min} = -\left[\left(s-m_{e'}^2-m_b^2\right)^2 - 4m_{e'}^2 m_b^2\right]/s$.

Fig.~\ref{fig:ReacDens} shows that the corresponding reaction density is by far out-of-equilibrium in the region of parameters of our interest, therefore the dark-electron kinetic equilibrium with the SM does not last longer than their chemical equilibrium.

\section{Renormalisation-scale dependence of the bounce action}\label{app:RGImprov}
%%%%%%%%%%%%%%%%%%%%%%%%%%%%%%%%%%%%%%%%%%%%%%%%%%%%%%%

In this appendix, we detail the derivation of Eq.~\eqref{eq:S3}. In particular, we justify the presence of the overall factor $Z_\phi^{3/2}$ in the expression of $S_3$, which has been overlooked in the previous literature. 

To illustrate our point, we begin our discussion by considering the classic Coleman-Weinberg model at $T=0$ \cite{Coleman:1973jx}, i.e.~the theory of massless scalar QED. In theories of radiative symmetry breaking, the effective action has to be used to derive the equations of motion for the bounce solution, i.e.~the instanton configuration describing the tunnelling process \cite{Weinberg:1992ds}. The effective action is a non-local object, however a derivative expansion can be carried out and, in the Euclidean, it reads
\be\label{eq:Seff}
    S_{\rm eff}(\phi) \approx \int d^4 x \left\{ V_{\rm eff}\left(\phi, c_i;\mu\right) + \frac{1}{2}Z_{\rm eff}\left(\phi, c_i;\mu\right) \partial_\nu \phi\partial^\nu \phi + \mathcal{O}\left(\partial^4\right)\right\}.
\ee
The first term in the expansion is the effective potential, while $Z_{\rm eff}$ is the coefficient of the two-derivative piece (not to be confused with the wavefunction renormalisation of the scalar field $Z_\phi$ in Eq.~\eqref{eq:Zphi}). While, in general, terms which are of higher order in the derivatives have a sizeable effect on the bounce, in the theory under study we can neglect them \cite{Weinberg:1992ds}, and truncate the derivative expansion at the quadratic order. The coefficients $c_i$ represents the generic couplings of the theory, which are solutions to the corresponding $\beta$ functions. We have made apparent the dependence on the couplings, as well as the explicit dependence on the renormalisation scale $\mu$, coming from one-loop terms. 

Requiring that $dS_{\rm eff}/d\log\mu =0$ gives \cite{Miransky:1994vk}
\be\label{eq:RGEInv}
\begin{aligned}
&V_{\rm eff}\left(\phi,c_{i};\mu\right)=V_{\rm eff}\left(\phi_0,c_{i,0};\mu_0\right), \\
&Z_{\rm eff}\left(\phi,c_i;\mu\right)=Z_{\rm eff}\left(\phi_0,c_{i,0};\mu_0\right) Z_\phi(\mu)\equiv \left[1+\delta Z_{\rm eff}^{(1)}\left(\phi_0,c_{i,0};\mu_0\right)\right]Z_\phi(\mu),
\end{aligned}
\ee
where $\delta Z_{\rm eff}^{(1)}$ represents the one-loop contribution to $Z_{\rm eff}$. However, its contribution to the bounce action is a two-loop effect \cite{Weinberg:1992ds}, and, hence, we neglect it hereafter by taking $Z_{\rm eff}\left(\phi,c_i;\mu\right)\approx Z_\phi(\mu)$.

At zero temperature, the bounce $\phi_b(r)$ is an O(4)-symmetric field configuration which is solution to the equation of motion 
\be
    Z_\phi\frac{d^2 \phi_{b}}{d r^2}+ Z_\phi\frac{3}{r}\frac{d\phi_{b}}{dr}=\frac{\partial V_{\rm eff}\left(\phi_{b},c_{i};\mu\right)}{\partial \phi_{b}}
\ee
with the boundary conditions
\be\label{eq:InCond}
\phi_{b}(r\rightarrow  \infty)= \phi_{t},\qquad \frac{d \phi_{b}(r)}{dr}\bigg\rvert_{r=0}=0,
\ee
where $r^2 = \tau^2 + x^2 + y^2+z^2$, and $\phi_{t}$ is the true minimum of the effective potential. Here, we work under the implicit assumption that the metastable minimum is at the origin $\phi_{f}=0$, and that $V_{\rm eff}\left(\phi_{f},c_{i};\mu\right)=0$. The O(4)-symmetric bounce describes quantum tunnelling at zero temperature.  

Assuming to have found $\phi_b$, the bounce action reads 
\be
    S_4\left(\phi_b\right)\approx 2\pi^2\int r^3 dr \left\{ \frac{1}{2} Z_\phi\left(\frac{d\phi_b}{dr}\right)^2 + V_{\rm eff}\left(\phi_b,c_i;\mu\right)\right\},
\ee
which is manifestly invariant under a change of $\mu$ (cfr.~Eq.~\eqref{eq:RGEInv}), up to two-loop-order corrections.

We suppose that the above argument extends to the scenario considered in the main body of the paper: a massless, real scalar field transforming in the adjoint of $\SU{2}$, two Weyl fermions in the fundamental, and a transition which is triggered by thermal effects. In particular, we assume that: 1) higher-derivative terms in the effective action are still subleading; 2) $\delta Z_{\rm eff}^{(1)}$ still represents a higher-order correction to the bounce action, even when it includes thermal corrections. Under such assumptions, we can derive the equation of motion for the O(3)-symmetric bounce describing thermal tunnelling from the finite-$T$ version of Eq.~\eqref{eq:Seff},
\be\label{eq:EoMS3}
    Z_\phi\frac{d^2 \phi_{b}}{d r^2}+ Z_\phi\frac{2}{r}\frac{d\phi_{b}}{dr}=\frac{\partial V_{\rm eff}\left(\phi_{b},c_{i};\mu,T\right)}{\partial \phi_{b}},
\ee
where the boundary conditions are given in Eq.~\eqref{eq:InCond}. The corresponding action is
\be
    S_3\left(\phi_b\right)\approx 4\pi\int r^2 dr \left\{ \frac{1}{2} Z_\phi\left(\frac{d\phi_b}{dr}\right)^2 + V_{\rm eff}\left(\phi_b,c_i;\mu,T\right)\right\}.
\ee
Notice that, now, the effective potential also depends on the temperature $T$, as it receives thermal corrections. 

In the high-temperature expansion, the thermal effective potential can be written as a polynomial in the field $\phi$ with coefficients that depend on the couplings as well as on the temperature
\be
\begin{aligned}
    V_{\rm eff}\left(\phi,c_i; \mu, T\right) &\approx \frac{m^2(c_i; T)}{2}\phi^2-\frac{\delta(c_i;T)}{3}\phi^3+\frac{\lambda(c_i;\mu,T)}{4}\phi^4\\
    &\equiv \frac{m_T^2}{2}\phi^2-\frac{\delta_T}{3}\phi^3+\frac{\lambda_T}{4}\phi^4.
\end{aligned}
\ee
We can then perform the following redefinitions: $\phi_b = \xi \varphi$, $r = L \rho$. By choosing $L = Z_\phi^{1/2}/m_T$ and $\xi = m_T^2/\delta_T$, Eq.~\eqref{eq:EoMS3} becomes
\be
    \frac{d^2 \varphi}{d\rho^2} + \frac{2}{\rho}\frac{d \varphi}{d \rho} = \frac{\partial \tilde{V}_{\rm eff}\left(\varphi, \kappa_T\right)}{\partial \varphi},
\ee
with 
\be
    \tilde{V}_{\rm eff}\left(\varphi, \kappa_T\right) = \frac{1}{2}\varphi^2 - \frac{1}{3}\varphi^3 + \frac{\kappa_T}{4}\varphi^4,\qquad \kappa_T \equiv \lambda_T \frac{m_T^2}{\delta_T^2}.
\ee
An analogous redefinition at the level of the bounce action gives
\be
    S_3(\varphi) = 4\pi Z_\phi^{3/2}\frac{m_T^3}{\delta_T^2}\int d\rho \rho^2 \left\{\frac{1}{2}\left(\frac{d\varphi}{d\rho}\right)^2 + \tilde{V}_{\rm eff}\left(\varphi, \kappa_T\right) \right\}\equiv 4\pi Z_{\phi}^{3/2}\frac{m_T^3}{\delta_T^2} \bar{S}_3(\kappa_T).
\ee
The equation of motion, as well as the reduced bounce action $\bar{S}_3$, are now functions of a single parameter $\kappa_T$. This allows for a numerical evaluation of $\bar{S}_3$ which can then be fitted, resulting in Eq.~\eqref{eq:S3} \cite{Adams:1993zs,Levi:2022bzt}.

\section{Gravitational wave signal}\label{app:GWs}
%%%%%%%%%%%%%%%%%%%%%%%%%%%%%%%%%%%%%%%%%%%%%%%%%%%%%%

sFOPTs are accompanied by a copious production of GWs, which appear as a stochastic background in the late-time Universe. The characteristic frequency and the intensity of such a signal both depend on the temperature of the phase transition, as well as on the latent heat released (see e.g.~Ref.~\cite{Athron:2023xlk} for a review, and Ref.~\cite{Maggiore:2018sht} for a textbook introduction). For instance, a strong transition occurring at a temperature around 100 MeV may explain the hint of detection in the nanohertz frequency region reported by the PTA collaborations \cite{EPTA:2023xxk ,NANOGrav:2023hvm}. As we will show in this section, phase transition scales relevant for monopole DM are typically associated to frequencies of the order of $\mathcal{O}(10^4)$ Hz, and they may be probed in the near future by the Einstein Telescope (ET) \cite{ET:2019dnz} and the Cosmic Explorer (CE) \cite{Reitze:2019iox}.\footnote{In this section, an instantaneous reheating after the phase transition is assumed. In general, an epoch of early matter domination preceding reheating is possible. This would modify the GW signal for modes re-entering the horizon during matter domination \cite{Hook:2020phx,Ellis:2020nnr,Gouttenoire:2023pxh,Racco:2025ons}, shifting it to lower frequencies. However, in the region of parameters of interest to us we do not observe the onset of matter domination (see the discussion in Sec.~\ref{sec:sFOPT}).}

During a FOPT, both bubble-wall collisions and the subsequent motion of the plasma fluid source GWs. Plasma motion manifests itself through the propagation of compression waves, which should eventually result into turbulence. However, numerical simulations were not able to observe the onset of the turbulent stage \cite{Caprini:2019egz}. Therefore, we assume that the only contribution to GW emission from the plasma motion comes from the propagation of sound waves. We express the fraction of energy density in GWs of frequency $f$ today as
\be\label{eq:OmGW}
\Omega_{\rm GW}(f)h^2 = \Omega_{\phi}(f)h^2 + \Omega_{{\rm sw}}(f)h^2,
\ee
where the first term is related to the GWs produced by the scalar field (bubble collisions), while the second one refers to the sound-wave contribution.

The dominant source of gravitational radiation in Eq.~\eqref{eq:OmGW} is determined by how the latent heat of the transition is shared between the scalar field kinetic energy and the bulk fluid motion. This, in turns, depends on the magnitude of the friction acting on the bubble walls and opposing their expansion.

\subsection{Plasma friction and energy budget}
After being nucleated, bubbles of the broken phase expand outwards, driven by the vacuum pressure $\Delta V$. At the same time, the interaction between the bubble wall and the surrounding plasma exerts a friction force oriented towards the inside of the bubble.

The leading-order contribution to the friction pressure is due to the plasma particles that acquire a mass in the broken phase \cite{Bodeker:2009qy,Azatov:2019png},
\be
    P_{\rm LO}(T) = \sum_i n_i c_i \frac{\Delta m_i^2 T^2}{24}\,.
\ee
Here, $n_i$ is the number of degrees freedom of the $i$-th particle species, $\Delta m_i^2$ represents its mass squared difference between the inside and the outside of the bubble, and $c_i =1 (1/2)$ for bosons (fermions). Since $\lambda \ll g^2$ in the case relevant for GW production (i.e.~for a sFOPT), and since we choose $y\ll g$ in our analysis of monopole production, in the case of interest to us the dominant contribution to $P_{\rm LO}$ is provided by the massive gauge bosons ${W'}^\pm$
\be
    P_{\rm LO}(T) \approx \frac{g^2 \eta^2}{4}T^2\,.
\ee
Bremsstrahlung emission of soft vector bosons by particles traversing the bubble wall provides the next-to-leading order correction to the friction pressure \cite{Bodeker:2017cim,Azatov:2019png}
\be
    P_{\rm NLO}(T) \approx \gamma_w \frac{g^3}{16\pi^2}\eta T^3,
\ee
where $\gamma_w = \left(1-v_w^2\right)^{-1/2}$ is the Lorentz boost factor of the wall. Remarkably, $P_{\rm NLO}$ grows with $\gamma_w$, thus preventing the bubble walls from reaching velocities larger than the terminal one
\be
    \bar\gamma_{w} = \frac{\Delta V - P_{\rm LO}(T_*)}{g^3\eta T_*^3/{16\pi^2}}\,,
\ee
for which $\Delta V = P_{\rm LO} + P_{\rm NLO}$. However, $P_{\rm NLO}$ only becomes relevant for large enough values of $\gamma_w$, i.e.~for bubbles that, in the absence of next-to-leading-order corrections to friction, would collide with a Lorentz factor $\gamma_{w,*} > \bar \gamma_w$, where \cite{Azatov:2019png}
\be\label{eq:gammawp}
\gamma_{w,*} \approx \frac{2}{3}\frac{R_*}{R_c^{\rm thin}}\left(1-\frac{P_{\rm LO}(T_*)}{\Delta V}\right)\,.
\ee
In Eq.~\eqref{eq:gammawp}, $R_c^{\rm thin}$ is the critical radius in the thin-wall approximation, which is given by
\be
    R_c^{\rm thin} = \frac{2}{3} \frac{Z_{\phi}^{1/2}}{1 - \frac{9}{2}\kappa(T_n)}\frac{1}{m_T(T_n)}\,,
\ee
with $Z_\phi$ the scalar field wavefunction renormalisation of Eq.~\eqref{eq:Zphi}, and $m_T$ and $\kappa_T$ the thermal effective potential coefficients defined in Eq.~\eqref{eq:ParamsT} and Eq.~\eqref{eq:S3}, respectively.
Notice that bubbles are typically produced with a thick-wall profile in sFOPTs. Nonetheless, Ref.~\cite{Azatov:2019png} found that, in the $\U{1}$ conformally-invariant gauge theory without fermions, the scalar field radial profile quickly evolves to a step-like function, characteristic of thin-wall bubbles, before bubble expansion is significant. We assume that this holds true also in our model, hence justifying Eq.~\eqref{eq:gammawp}.

In transitions with $\gamma_{w,*} < \bar \gamma_w$, $P_{\rm NLO}$ is never relevant and bubble walls exhibit a runaway behavior, as they keep accelerating till the moment of percolation. In the runaway regime most of the latent heat is transferred to bubble wall motion, and the main source of GWs is represented by wall collisions. On the contrary, if $\gamma_{w,*} \geq \bar \gamma_w$, bubbles reach a terminal velocity and most of the vacuum pressure goes into bulk fluid motion. Following Ref.~\cite{Azatov:2019png}, we define the \textit{scalar field efficiency factor}, i.e.~the fraction of energy going into wall motion, as 
\be\label{eq:kphi}
    \kappa_\phi \approx \begin{cases}
        \frac{\bar\gamma_w}{\gamma_{w,*}}\left(1-\frac{\alpha_\infty}{\alpha}\right), \quad \bar\gamma_w\leq\gamma_{w,*}\\
    1-\frac{\alpha_\infty}{\alpha},\quad \bar\gamma_w>\gamma_{w,*}     
    \end{cases}
\ee
where $\alpha$ is defined by Eq.~\eqref{eq:alpha} and $\alpha_\infty \equiv P_{\rm LO}/\rho_r(T_p)$. By energy conservation, the \textit{plasma efficiency factor} is $\kappa_{\rm pl}\approx 1-\kappa_\phi$.

\begin{figure}[h!]
\begin{center}
\includegraphics[width=.48\textwidth]{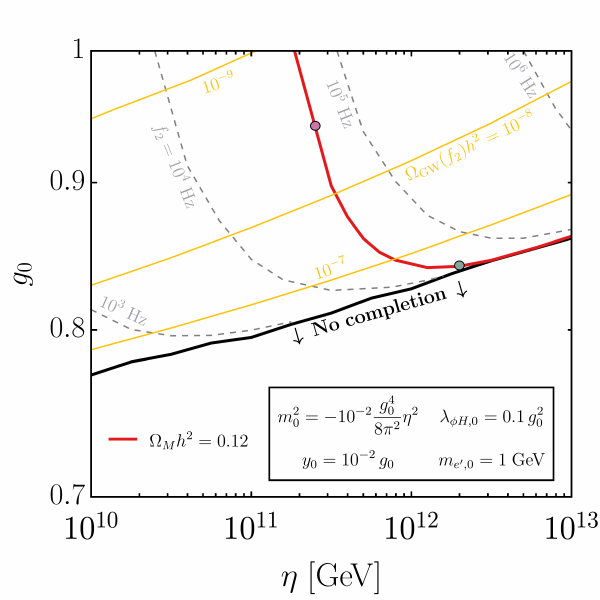}\quad \includegraphics[width=.48\textwidth]{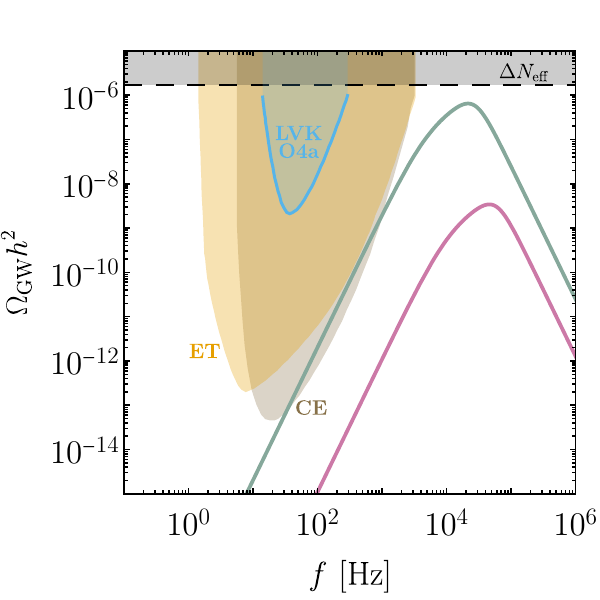}
    \caption{\textit{Left panel.} In yellow, we show isocontours of the amplitude of the GW signal produced during the phase transition, evaluated at the frequency $f_2$ defined in Eq.~\eqref{eq:f1f2}. Along the dashed gray lines, the characteristic frequency $f_2$ is constant. Sound waves represent the dominant GW source in the whole parameter space. The choice of parameters is the same as in the left panel of Fig.~\ref{fig:MonoFermsFOPT}. For comparison, we plot in red the curve along which monopoles comprise the totality of DM. We pick two benchmark points on this line and show the related GW signal in the \textit{right panel}. Below the solid black line, the transition never completes. \textit{Right panel.} The green and the pink lines represent the GW signal emitted during the phase transition, computed for the corresponding parameter-space points in the \textit{left panel}. The projected sensitivity curve for the Einstein Telescope \cite{ET:2019dnz} (Cosmic Explorer \cite{Reitze:2019iox}), as derived in Ref.~\cite{NANOGrav:2023hvm} assuming a signal-to-noise ratio equal to five and an observing time of ten years, is shown in light orange (brown). The exclusion region set by the O4a run of the LIGO-Virgo-KAGRA collaboration \cite{LIGOScientific:2025kry} is shown in light blue. The dashed black line represents the Planck \cite{Planck:2018vyg} $\Delta N_{\rm eff}$ constraint on the relic density of GWs \cite{Smith:2006nka, LISACosmologyWorkingGroup:2022jok}.}
\label{fig:GWs}  
\end{center}
\end{figure}

In Fig.~\ref{fig:GWs}, we study the GW signal produced by a sFOPT for the same parameters as in Fig.~\ref{fig:MonoFermsFOPT}. We find that $\gamma_{w,*} > \bar \gamma_w$, therefore sound waves are the dominant source of GW production, with $\kappa_{\rm pl} \gg \kappa_\phi$. In the left-hand panel of Fig.~\ref{fig:GWs} we show isocontours of $f_2$ and $\Omega_{\rm GW}(f_2)h^2$ in the plane $(\eta,\,g_0)$, where the relevant frequency $f_2$ will be defined in Eq.~\eqref{eq:f1f2}. Along the line where monopoles comprise the totality of DM,  we choose two benchmark points and show the corresponding GW signal in the right-hand panel of Fig.~\ref{fig:GWs}.

\subsection{Bubble collisions}

We follow Ref.~\cite{Caprini:2024hue} and model the shape of the GW spectrum sourced by bubble collisions with a broken power law
\be\label{eq:S}
    \mathcal{S}(f) = \left[\frac{a}{\left(\frac{\hat f_{\phi}}{f}\right)^{b} + \left(\frac{f}{\hat f_{\phi}}\right)^{c}}\right]^d,
\ee
where $\hat f_{\phi}$ is the peak frequency. Lattice simulations of strong transitions with $\alpha \gg 1$ give $a\approx 2$, $b \approx c \approx 0.6$, $d\approx 4$ \cite{Lewicki:2022pdb}. For such values, the small-frequency limit of the spectral shape in Eq.~\eqref{eq:S} does not exhibit the typical $f^3$ scaling that one would expect from causality in a radiation-dominated universe \cite{Caprini:2009fx}. The reason is that the numerical simulations in Ref.~\cite{Lewicki:2022pdb} do not take into account the expansion of the Universe. We therefore impose by hand the correct behaviour for frequencies smaller than the Hubble radius at the moment of the transition, as done e.g.~in Ref.~\cite{Gouttenoire:2023pxh}. All in all, the GW spectrum reads
\be
    \Omega_{\phi}(f)h^2  \approx \hat{\Omega}_\phi h^2 \mathcal{S}(f)\frac{\left(\frac{f}{f_H}\right)^{3-bd}}{1+\left(\frac{f}{f_H}\right)^{3-bd}}
    \approx\hat{\Omega}_\phi h^2 \left[\frac{2}{\left(\frac{\hat f_{\phi}}{f}\right)^{0.6} + \left(\frac{f}{\hat f_{\phi}}\right)^{0.6}}\right]^4 \frac{\left(\frac{f}{f_H}\right)^{0.6}}{1+\left(\frac{f}{f_H}\right)^{0.6}}\,,
\ee
where $f_H$ is the frequency related to the Hubble radius at the moment of the phase transition,
\be
    f_H = 2.69\times 10^3\,{\rm Hz} \left(\frac{g_{\star,{\rm reh}}}{115}\right)^{1/6}\left(\frac{T_{\rm reh}}{10^{11}\,{\rm GeV}}\right)\,.
\ee
The peak frequency and the peak amplitude are given respectively by
\cite{Lewicki:2022pdb}
\be\label{eq:peakScalar}
\begin{aligned}
    &\hat{f}_{\phi} \approx 8.23 \times 10^3 \,{\rm Hz} \left(\frac{g_{\star,{\rm reh}}}{115}\right)^{1/6} \left(\frac{T_{\rm reh}}{10^{11}\,{\rm GeV}}\right)\left(H_{\rm reh} R_*\right)^{-1},\\
    &\hat{\Omega}_{\phi}h^2 \approx 3.96\times 10^{-8} \left(\frac{115}{g_{\star,{\rm reh}}}\right)^{1/3} \left(\frac{\kappa_\phi \alpha}{1+\alpha}\right)^2 \left(H_{\rm reh} R_*\right)^2 .
\end{aligned}
\ee
As expected, the amplitude of the signal depends on the efficiency factor $\kappa_\phi$ given in Eq.~\eqref{eq:kphi}, i.e.~on the amount of vacuum energy that is transferred to bubble-wall motion.

\subsection{Sound waves}
%%%%%%%%%%%%%%%%%%%%%%%%%%

The signal generated by sound waves depends on both the typical size of colliding bubbles and on the thickness of the fluid shells, each associated with a characteristic frequency, $f_1$ and $f_2$ respectively. Therefore, the GW spectrum can be modelled using a broken power law fitted to numerical simulations \cite{Caprini:2024hue, Jinno:2022mie}
\be
    \Omega_{\rm sw}(f)h^2 = \Omega_{{\rm int}}h^2 N \left(\frac{f}{f_{1}}\right)^3 \left\{ \left[1+\left(\frac{f}{f_{1}}\right)^2\right] \left[1+\left(\frac{f}{f_{2}}\right)^4\right] \right\}^{-1},
\ee
where $N$ is a normalisation factor such that $\int_{-\infty}^{+\infty} d \log f\;\Omega_{{\rm sw}}(f)h^2=\Omega_{{\rm int}}h^2$. The characteristic frequencies are \cite{Jinno:2022mie}
\be
\begin{aligned}\label{eq:f1f2}
    f_{1} &\approx   3.38\times 10^3\,{\rm Hz} \left(\frac{g_{\star,{\rm reh}}}{115}\right)^{1/6} \left(\frac{T_{\rm reh}}{10^{11}\,{\rm GeV}}\right)\left(H_{\rm reh}R_*\right)^{-1},\\
    f_{2} &\approx 8.45\times 10^3\,{\rm Hz} \left(\frac{g_{\star,{\rm reh}}}{115}\right)^{1/6} \left(\frac{T_{\rm reh}}{10^{11}\,{\rm GeV}}\right) \left(\Delta_w H_{\rm reh}R_*\right)^{-1},
\end{aligned}
\ee
where $\Delta_w$ is the dimensionless shell thickness, which for relativistic wall velocities is $\Delta_w \approx 0.42$ \cite{Caprini:2024hue}. The integrated energy density reads \cite{Jinno:2022mie}
\be
    \Omega_{{\rm int}}h^2 \approx 1.71\times 10^{-6}\left(\frac{115}{g_{\star,{\rm reh}}}\right)^{1/3}\left(\frac{\kappa_{\rm pl} \alpha}{1+\alpha}\right)^2 \min \left[\frac{H_{\rm reh} R_*}{\bar{v}_f},1\right] \left(H_{\rm reh} R_*\right) ,
\ee
with $\bar v_f = \frac{3}{4} \frac{\kappa_{\rm pl} \alpha}{1+\alpha}$ the average fluid velocity. 

Isocontours of $f_2$ and $\Omega_{\rm GW}(f_2)h^2$ in the plane $(\eta,g_0)$ are shown in the left-hand panel of Fig.~\ref{fig:GWs}, alongside the monopole relic abundance (the spectrum peaks around $f_2$). The monopole line is typically associated to large amplitudes of the GW  signal and characteristic frequencies of tens of kilohertz. In the right-hand panel in Fig.~\ref{fig:GWs}, we show the GW signal for two benchmark points where monopoles comprise the totality of DM, together with current bounds and projected sensitivity curves from various experiments. While both $f_1$ and $f_2$ are too large to be probed in the future by interferometers, the amplitude of the signal is large enough for the IR tail to fall inside the sensitivity regions of the ET and the CE, for $\eta \simeq 2\times 10^{12}\,{\rm GeV}$ and $g_0 \simeq 0.85$.

\bibliographystyle{hieeetr}
\bibliography{MonoDM.bib}

\end{document}